\documentclass[aps,prb,reprint,superscriptaddress]{revtex4-2}
\usepackage{hyperref}
\usepackage{xcolor}
\usepackage{graphicx}
\usepackage{subcaption}
\usepackage{amsmath}
\usepackage{physics}
\usepackage{dsfont}

\graphicspath{{./figures/}}

\begin{document}
	
	\title{Two parameter scaling in the crossover from symmetry class BDI to AI}
	
	\date{\today}
	
	\author{Saumitran Kasturirangan}
	\affiliation{School of Physics and Astronomy, University of Minnesota, Minneapolis, Minnesota 55455, USA}
	\author{Alex Kamenev}
	\affiliation{School of Physics and Astronomy, University of Minnesota, Minneapolis, Minnesota 55455, USA}
	\affiliation{William I. Fine Theoretical Physics Institute, University of Minnesota, Minneapolis, Minnesota 55455, USA}
	\author{Fiona J. Burnell}
	\affiliation{School of Physics and Astronomy, University of Minnesota, Minneapolis, Minnesota 55455, USA}

	\begin{abstract}
		The transport statistics of the 1D chain and metallic armchair graphene nanoribbons with hopping disorder are studied, with a focus on understanding the cross-over between the zero-energy critical point and the localized regime at larger energy. In this cross-over region, transport is found to be described by a 2-parameter scaling with the ratio $s$ of system size to mean-free-path, and the product $r$ of energy and scattering time.  This 2-parameter scaling shows excellent data collapse across a wide a variety of system sizes, energies, and disorder strengths.  The numerically obtained transport distributions in this regime are found to be well-described by a Nakagami distribution, whose form is controlled up to an overall scaling by the ratio $s / |\ln r|^2$.  For sufficiently small values of this parameter, transport appears virtually identical to that of the zero-energy critical point, while at large values, a Gaussian distribution corresponding to exponential localization is recovered.  For intermediate values, the distribution interpolates smoothly between these two limits.

	\end{abstract}
	
	\maketitle
	
\section{Introduction}

In the presence of disorder, most 1D systems are Anderson insulators \cite{AndersonLoc, LeeRamakrishnanReview, KramerMackinnonReview, EversMirlinReview}, with electronic wavefunctions that are exponentially localized in space.  This phenomenon is most easily understood by the classic ideas of the scaling theory of localization \cite{ScalingTheoryLoc, GorkovLarkinScaling}, which reveals that disorder is relevant in 1D (and marginally relevant in 2D) such that low-dimensional systems generically localize in the presence of arbitrarily weak disorder.  Localization has many dramatic experimental signatures -- notably, the conductance decreases exponentially with the system size, such that disordered 1D metals are generically insulating.

However, there are exceptions to this rule.  A famous example of this, first studied by Dyson \cite{DysonDOS}, is the 1-dimensional chain with only hopping disorder. Near zero energy, this system has several unusual spectral properties. First, as noted by Dyson \cite{DysonDOS},  the low-energy density of states diverges with energy $\epsilon$ as $1/( |\epsilon \ln^3 |\epsilon|) $.  The typical localization length also diverges logarithmically with the energy in the limit of zero energy \cite{Theodorou1DChain, EggarterLocDivergence, ZimanLocDivergence, DFisherIsingChains, McKenzieQPT,Kamenev2014},  and precisely at zero energy the conductance does not decay exponentially with system size, even for long wires, indicating that exponential localization is never reached. Simultaneously, the exponentially localized wave functions give way to electronic eigenstates with a multifractal character \cite{BalentsFisher,Kamenev2016}.

The unusual features of the random hopping model at zero energy are reflected in its transport properties. These can be studied using a Fokker-Plank (FP) equation, which describes the evolution of the probability distribution of transport with system size. The original equation of this type, derived by Dorokhov \cite{Dorokhov82} and Mello, Pereyra, and Kumar \cite{MPK} for the Wigner-Dyson symmetry classes, predicts a conductance that falls off exponentially with the system size for long enough systems, and hence a finite localization length. More generally, however, a different FP equation can be derived for each of the 10 different Altland-Zirnbauer symmetry classes\cite{AltlandZirnbauer, MelloStone,MacedoChalker,Deloc1DChains,MudryRandomFlux,LocalizationDelocalizationDirtySC,Kamenev2015}. The 1D chain with real hopping disorder falls into symmetry class BDI, for which solutions to the FP equation at zero energy are not exponentially localized for an odd number of channels\cite{MudryRandomFlux,Deloc1DChains}. This is also the case for the other chiral classes AIII and CII when there are an odd number of channels, as well as the disordered superconducting classes D and DIII for any number of channels \cite{LocalizationDelocalizationDirtySC}.

The diverging localization length suggests that Dyson's random hopping model is fine-tuned to an underlying (disorder-induced) critical point at zero energy. Indeed,  a uniform staggering of the hopping preserves all symmetries but generically tunes the system away from criticality \cite{Deloc1DChains,MotrunichDamleHuse}; the divergences of localization length and density of states are thus features of the underlying critical point, rather than of the symmetry class in general. (In multi-mode wires there is in fact a 2-parameter family of critical points, see \cite{BrouwerChiralFP}). Moreover, Gruzberg et. al. \cite{GruzbergSuperuniversality} pointed out that the higher-order terms neglected in a standard FP treatment tune systems in all symmetry classes away from criticality, explaining the apparent generality of this critical behavior observed in earlier works such as \cite{LocalizationDelocalizationDirtySC}.

Nevertheless, the random hopping  critical point is relevant to understanding transport in a wide array of 1D systems, including multi-mode quasi-1D wires with random hopping, and disordered superconductors \cite{Deloc1DChains, BrouwerDOS, TitovDOS, BrouwerDirtySC}.   Indeed, this critical point is not only universal, in the sense of being independent of details of the disorder potential or the number of channels, but in fact super-universal, in the sense that the conductance statistics are universal across critical points that appear in five distinct symmetry classes \cite{GruzbergSuperuniversality}.

Interestingly, although divergences in the localization length and density of states near this critical point have been extensively studied, the full transport statistics near the critical point are not fully understood.  One underlying challenge here is that the FP equation corresponding to symmetry class BDI, describes transport of the 1D disordered chain only in the limit that charges are injected at precisely zero energy, where the chiral symmetry constrains transport. The resulting transport distributions are thus rigid for a given symmetry class, describing scaling only with  the ratio $s = L/l$ of length to mean-free-path. Consequently, they cannot capture the cross-over from the critical transport statistics at zero energy to the localized regime that occurs at higher energies.

One attempt to capture transport in this cross-over regime was made by Ryu et. al. \cite{RyuCrossover}, who derived FP-like equations to describe the finite energy cross-over regime.  These equations were obtained by relaxing certain assumptions made in the standard FP derivations, and lead to a 2-parameter scaling, where the second parameter $\tilde{r} = \epsilon / V^2$ is proportional to the energy $\epsilon$, and $V$ is the disorder strength.  However, the resulting equations cannot be solved analytically without making further approximations that are difficult to justify a priori.

The present work takes a different approach, using a numerical study of transport in the cross-over regime to directly access the relevant transport statistics. The numerics are carried out on two models tuned to the BDI zero-energy critical point: the 1D chain with random hopping disorder, and metallic armchair graphene nanoribbons at low-energies, where there is only one propagating mode.

This approach reveals several qualitative aspects of transport in the cross-over regime which have not been discussed in the literature to date. First, the probability distribution of transport quantities, such as the conductance, is a universal function of two dimensionless parameters that are determined by the system size, energy, and disorder strength. These parameters are: the ratio $s = L/l$ of length to mean-free-path, and the product $r = \epsilon \tau$ of energy and relaxation-time.  Such a 2-parameter scaling is also consistent with the generalized FP equations of ref. \cite{RyuCrossover}.  However, the scaling parameter that yields the best data collapse at larger disorder strengths is $r$, rather than $\tilde{r}$.  (The two agree at weak disorder.)

\begin{figure}
	\centering
	\includegraphics[width=\linewidth]{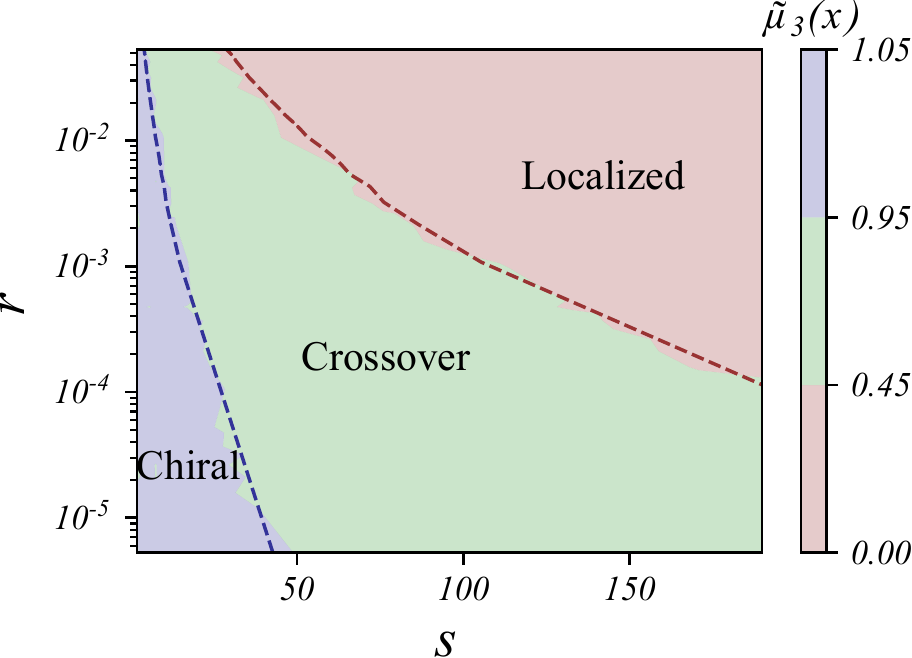}
	\caption{Evolution of the probability distribution $P(x)$ as a function of the two parameters $s = \frac{L}{l}$ and $r = \epsilon\tau$. In the blue region, for sufficiently small $s/|\ln r|^2$, the distribution is indistinguishable from that of the Chiral class. In the red region, when $s/|\ln r|^2$ is large, one enters the localized regime at any finite $r$, where the typical conductance decays exponentially with length. Here, the distribution $P(x)$ is approximately  Gaussian, with a mean exceeding its standard deviation. In the green region, the distribution crosses over  smoothly between these two limits. The color map shows the skewness $\tilde{\mu}_3(x)$, which effectively distinguishes between these regimes. For more details, see  section \ref{Sec:Results}.}
	\label{fig:phase_portrait}
\end{figure}

Second, as a function of the two parameters $(r,s)$, transport is found to be in one of three regimes, as shown in Fig. \ref{fig:phase_portrait}.  For any finite $r$, as $s \rightarrow \infty$ transport enters a localized regime with the typical conductance decaying exponentially with system size.  Conversely, when $s \lesssim  | \ln r |^2$, the transport distributions are essentially identical to the distribution of the chiral class at zero energy, which can be understood as a consequence of the fact that the wire is not long enough to reveal its finite localization length.  These two limits are separated by a cross-over regime, where the shape of the probability distribution changes smoothly with $(r,s)$.

Third, throughout the cross-over regime, the ratio $s/|\ln r|^2$, which describes the ratio of the system size to average localization length, is found to control the overall shape of the transport distribution, up to an overall re-scaling. For $s/|\ln r|^2 \lesssim 0.28$, the distribution is numerically indistinguishable from that of the critical point at zero energy, while for $ s /|\ln r|^2 \gtrsim 2$, transport appears localized.

The rest of this paper is structured as follows. Sec. \ref{Sec:Background} reviews the usual single parameter scaling paradigm, the FP equations governing transport, and the properties of their solutions. These can be thought of as the limiting cases of the cross-over problem under consideration here. The present study focuses on two models: the 1D chain and arm-chair graphene ribbon. These are presented in Sec. \ref{Sec:Models}, which also describes symmetries of the models and the methods used to compute the conductance numerically. The main results of this work are discussed in Sec. \ref{Sec:Results}. The two-parameter scaling of transport statistics is established using data-collapse, followed by an analysis of the relevant probability distributions, and comparisons to prior results from the literature.  Sec. \ref{Sec:Conc} presents the conclusions with some remarks on the scope for future work.

\section{Background} \label{Sec:Background}

This section covers some of the background from previous works that is required to understand the results of the present paper. The conventional FP approach for studying the evolution of the probability distribution of transport with the system size is reviewed. The solutions of the FP equations for the relevant symmetry classes AI and BDI are discussed. Note that the current work studies the cross-over of the transport statistics between the BDI symmetry class at the zero energy critical point and the AI symmetry class expected at large energies. Finally, the inherent inadequacy of the conventional FP approach, as well as the associated single parameter scaling, in describing this cross-over is presented.

The long-distance transport properties of a generic disordered, non-interacting fermionic system depend qualitatively on its symmetries. In the absence of lattice symmetry, there are 10 distinct symmetry classes, characterized by the presence or absence of three generic symmetries: time reversal ($T$), particle-hole ($C$), and chiral symmetry ($S$) \cite{AltlandZirnbauer}.  This work is focused on Hamiltonians in one of two symmetry classes: AI (with $T=+1$) and BDI (with $T=C=+1$ and $S=1$).  AI is also known as the Orthogonal class, and is one of the original Wigner-Dyson classes.  BDI has Chiral symmetry ($S=1$), which implies that there is a unitary operator that \emph{anti-commutes} with the (first-quantized) Hamiltonian (see \cite{RyuTenFoldWay}).  In the presence of Chiral symmetry, the spectrum is invariant under reflection about zero energy. Therefore, zero energy is a special point in the spectrum. The implications of this on electronic transport are discussed below.

One of the common ways to study transport in quasi-1D wires is to use the S-matrix that relates the amplitudes of incoming and outgoing modes at a given energy:
\begin{equation}
	\begin{pmatrix}
		\psi^{o}_L \\
		\psi^{o}_R
	\end{pmatrix} = S 
	\begin{pmatrix}
		\psi^{i}_L \\
		\psi^{i}_R
	\end{pmatrix} \equiv 
	\begin{pmatrix}
		\mathcal{R} & \mathcal{T}' \\
		\mathcal{T} & \mathcal{R}'
	\end{pmatrix}
	\begin{pmatrix}
		\psi^{i}_L \\
		\psi^{i}_R
	\end{pmatrix}.
\end{equation}
The labels $o (i)$ and $L (R)$ correspond to the outgoing (incoming) modes on the left (right) end of the wire. The quantities $\mathcal{R} (\mathcal{R}')$ and $\mathcal{T} (\mathcal{T}')$ are the reflection and transmission coefficients. For an $N$ channel system, these are $N \times N$ matrices. The conductance, in units of $e^2/h$, is then given by:
\begin{equation}
	g = \Tr\left(\mathcal{T} \mathcal{T}^\dagger\right) = \sum_{i=1}^{N} T_i,
\end{equation}
where $\{T_1, T_2, \dots , T_N\}$ are the eigenvalues of $\mathcal{T} \mathcal{T}^\dagger$. It is often convenient to re-write the transmission eigenvalues $T_i$ as:
\begin{equation}
	T_i = \frac{1}{\cosh[2](x_i)}.
	\label{eqn:x_definition}
\end{equation}
The values $x_i$ are related to the eigenvalues of the transfer matrix $M$ that relates the amplitude of modes on the left and right sides of the wire. The eigenvalues of $M M^\dagger$ come in pairs $\exp(\pm 2 x_i)$ \cite{BeenakkerRMT, GruzbergSuperuniversality}. These are often referred to as the radial coordinates of the transfer matrix. For the purposes of the present work, it is sufficient to consider systems with a single propagating channel (after eliminating degeneracies), i.e. $N=1$. In this case, there is only one eigenvalue, so the index $i$ is dropped henceforth.

By considering how the S-matrix changes upon adding a thin slice of disordered wire\cite{Anderson80}, a differential equation for the evolution of the probability distribution of transmission coefficients as a function of the wire's length can be obtained \cite{Dorokhov82,MPK} (for a review of these methods, see \cite{BeenakkerRMT, BrouwerFPReview}). 
The resulting Fokker-Plank, or DMPK, equation for a one-channel system is given by:
\begin{equation} \label{eq:FP}
	\pdv{P(x; s)}{s} = \frac{1}{2 \gamma}\left[ \pdv{}{x} \left( \pdv{P}{x} - P \pdv{\ln(J(x))}{x} \right) \right]
\end{equation}
Here $P(x; s)$ is the probability distribution of $x$, which is related to the transmission eigenvalue via equation \ref{eqn:x_definition}, and  
\begin{equation}
	s = \frac{L}{l}
\end{equation}
is the ratio of length to mean-free path. 

Both current conservation and the symmetries $T,C$, and $S$ (when present) impose constraints on the S-matrix (or transfer matrix). For instance, time reversal symmetry enforces that the S-matrix at a given energy is symmetric:
\begin{equation}
	S(\epsilon) = \left(S(\epsilon)\right)^T.
\end{equation}
However, chiral symmetry relates the S-matrix at energy $\epsilon$ to that at $-\epsilon$:
\begin{equation}
	S(\epsilon) = \left( S(-\epsilon) \right)^\dagger.
\end{equation}
Therefore, chiral symmetry constrains the S-matrix to be Hermitian only at precisely zero energy. 

Because the $S$-matrix in different symmetry classes satisfies different constraints, the quantities $\gamma$, $J(x)$, and the domain of $x$ are specific to a symmetry class and are given by:\cite{GruzbergSuperuniversality, BrouwerFPReview}
\begin{align*}
	\mathrm{AI:} &		&         0   &\leq  x < \infty;	&  \gamma  &= 2 ;   &  J(x) &= \sinh(2x) \\
	\mathrm{BDI:}&		&	-\infty   &< x < \infty;		&  \gamma  &= 1 ;   &  J(x) &= 1 
\end{align*}
This leads to different FP equations governing transport in AI and BDI.  We emphasize, however, that the FP equation corresponding to the BDI symmetry class is only applicable at strictly zero energy, where chiral symmetry requires the $S$ matrix to be Hermitian. 
One can also understand this by thinking of calculating transport at non-zero energy as analogous to introducing a finite chemical potential that breaks the chiral symmetry of the BDI symmetry class.

Assuming that $P(x; 0) = \delta(x)$, the solution for class AI is given by \cite{AbrikisovSolution}:
\begin{equation}
	\begin{split}
		P(x; s) &= \frac{1}{\sqrt{\pi}} \left( \frac{2}{s} \right)^{\frac{3}{2}} \sinh(2x) e^{-\frac{s}{4}}  \\ & \times \int_{x}^{\infty} \mathrm{d}y \, \frac{y e^{-\frac{y^2}{s}}}{\sqrt{\cosh(2y) - \cosh(2x)}}
	\end{split}
	\label{eqn:ai_prob}
\end{equation}
Similarly for class BDI, one obtains (for $x \in \mathds{R}$) \cite{BrouwerDirtySC, RyuCrossover}:
\begin{equation}
	P(x; s) = \frac{1}{\sqrt{2\pi s}} e^{-\frac{x^2}{2s}}
	\label{eqn:bdi_prob}
\end{equation}
Based on the FP equation and its solutions, one finds that the entire probability distribution (therefore all its moments) for a given symmetry class is completely specified by a single parameter $s$. This is sometimes referred to as one-parameter scaling.

It is now helpful to analyze the properties of the solutions to the FP equations in \ref{eqn:ai_prob} and \ref{eqn:bdi_prob} in more detail. According to the scaling theory of localization \cite{ScalingTheoryLoc}, a conventional 1D system is exponentially localized for arbitrarily weak disorder with:
\begin{equation}
	-\expval{\ln\left(g\right)} = 2\frac{L}{\xi_{typ}}, \quad -\ln\left(\expval{g}\right) = 2\frac{L}{\xi_{avg}}
	\label{eqn:xi}
\end{equation}
for long system sizes, where $\xi_{typ}(\xi_{avg})$ is the typical (average) localization length. Indeed this is what one finds for class AI, where:
\begin{equation}
	-\expval{\ln(g)} = s,
	\label{eqn:lng_ai}
\end{equation}
for any system size. Therefore $\xi_{typ} = 2l$ in this case.  It can also be shown that $\xi_{avg}=8l$ \cite{BeenakkerRMT}.  The latter becomes apparent at large system sizes $s \gg 1$, where the distribution of equation \ref{eqn:ai_prob} becomes a Gaussian with both mean and variance of $\frac{s}{2}$.

However, for the solution of class BDI eqn. \ref{eqn:bdi_prob}, exponential localization is never reached. Even in the asymptotic limit of $s \gg 1$, using equation \ref{eqn:bdi_prob} one finds:
\begin{equation}
	-\expval{\ln(g)} \approx 2\sqrt{\frac{2s}{\pi}}.
	\label{eqn:lng_bdi}
\end{equation}
The typical localization length scales as $\xi_{\text{typ}} \sim  \sqrt{Ll}$ in this case, and becomes infinite in the limit of large system size.  Similarly $\xi_{avg} \sim L/\log s$, and diverges with $L$. It must be noted that this does not correspond to a perfectly conducting system, but one with a broad transport distribution. In fact, based on equation \ref{eqn:bdi_prob}, one finds that $\ln(g)$ is not self-averaging in the BDI symmetry class, even for arbitrarily long system sizes. This is the disorder induced critical point in 1D \cite{LocalizationDelocalizationDirtySC,MotrunichDamleHuse,Kamenev2016}.

The existence of the critical point at zero energy is intimately connected to topology in non-interacting systems. In symmetry class BDI, it is possible to have topologically non-trivial gapped phases in 1D. These are classified by an integer-valued topological invariant of the band structure \cite{RyuTenFoldWay}, the winding-number. A disorder-free 1D metal in class BDI can be viewed as the critical point separating the topologically trivial phase with zero winding from one of the non-trivial phases with winding $\pm 1$. One can then imagine that adding disorder creates regions in each of these phases. On the domain walls between these disorder-induced topologically distinct regions, there exist low-energy states. It is these low-energy states which lead to the divergent density of states and localization length noted previously \cite{DysonDOS, TitovDOS, MotrunichDamleHuse}.

As shown above, the FP equation describes a 1-parameter scaling that changes abruptly from one symmetry class to another.  This poses a conundrum for scattering at non-zero energy in class BDI, where chiral symmetry does not impose any constraints on the S-matrix, and the relevant FP equation is technically in symmetry class AI. It follows that at sufficiently low energies, eqn. \ref{eq:FP} fails to describe transport distributions in such systems, which must cross over smoothly from that of class BDI at zero energy to that of class AI for sufficiently large energies (compared to the inverse relaxation-time, say). In this regard, energy can be thought of as a parameter that moves us away from criticality.  (As noted previously, zero-energy is not always a disorder-induced critical point, as symmetry preserving terms such as a staggered hopping can be added to tune the system away from criticality \cite{MotrunichDamleHuse, GruzbergSuperuniversality}; such terms are not considered here.)  A minimal description of the transport statistics in this finite energy cross-over from class BDI to AI thus requires one additional parameter, which clearly must be related to the energy; here we argue that the appropriate parameter is $r = \epsilon \tau$, where $\tau$ is the scattering time.  Note that a breakdown of one-parameter scaling also occurs in the standard Anderson localization problem near zero-energy and the band edges \cite{TitovShortRangeDisorder, TessieriBandCenterAnomaly, KandBandCenterAnomaly}; this is distinct from the phenomenon we discuss here.

Ref. \cite{RyuCrossover} suggested that the FP equation fails to describe  the finite enregy cross-over regime between symmetry class BDI and AI because the magnitude and phase of the reflection coefficient do not decouple. Specifically, writing the reflection coefficient as
\begin{equation}
	\mathcal{R} = \sqrt{R} e^{i \phi},
\end{equation}
where $\sqrt{R} = \tanh(x)$ is the amplitude and $\phi$ is the phase. The usual FP equation assumes that the probability distribution for the phase $Q(\phi)$ is stationary with respect to $s$, independent of $x$, and follows:
\begin{equation}
	Q(\phi) =
	\begin{cases}
		\frac{1}{2\pi}, & \text{AI} \\
		\frac{1}{2}[\delta(0) + \delta(\pi) ], & \text{BDI}
	\end{cases}
\end{equation}
Ref. \cite{RyuCrossover} derived FP equations which incorporate the flow of $\phi$ with $s$, and proposed approximate solutions to these in order to describe the scaling regime.   The drawback of this approach is that the resulting differential equations cannot be solved exactly; we return to this point in Sec. \ref{Sec:Results}, where the results of this work are compared to those of ref. \cite{RyuCrossover}.

\section{Model and Methods} \label{Sec:Models}

In this section, the models used to study the cross-over of transport from symmetry class BDI to AI are introduced. These are the 1D chain and metallic arm-chair graphene ribbon. The symmetries of the disordered models used to classify the symmetry class of these models is then discussed. Finally, a brief description of the methods used to compute transport numerically is given.

\begin{figure}
	\centering
	\begin{subfigure}{0.48\linewidth}
		\centering
		\includegraphics[width=\linewidth]{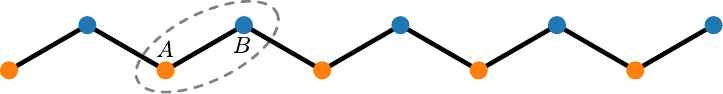}
		\caption{}
	\end{subfigure}
	\hfill
	\begin{subfigure}{0.48\linewidth}
		\centering
		\includegraphics[width=\linewidth]{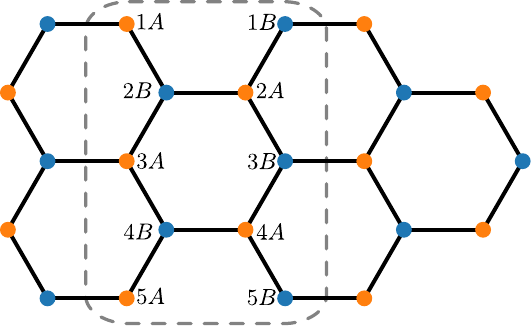} 
		\caption{}
	\end{subfigure}
	\caption{Examples of the systems studied here. The 1D chain is shown in figure (a) and the arm-chain ribbon of width $W=5$ in figure (b). Differently colored sites correspond to different sub-lattices.  The dashed line in  panel (a) (panel (b))  encloses a unit cell of the 1D chain (arm-chair ribbon). Throughout this work, the lattice constant is taken to be unity. The labelling of sites for the arm-chair ribbon is shown in figure (b), where the number corresponds to the vertical position of the site within the unit cell and the letter corresponds to the sub-lattice.}
	\label{fig:lat}
\end{figure}

To study the cross-over and establish its properties, two tight binding systems are considered. The first is a 1D chain and the second is a graphene nanoribbon with arm-chair boundaries. These are shown in figure \ref{fig:lat}. The Hamiltonian of the 1D chain in the absence of disorder is given by:
\begin{equation}
	H = \sum_i t (c^{\dagger}_{i, B} c_{i, A} + c^\dagger_{i+1, A} c_{i, B}) + \text{h.c.},
	\label{eqn:ham_1d}
\end{equation}
where $c_{i, A}$ ($c_{i, B}$) are annihilation operators for electrons on the unit cell labeled by the index $i$ at the sub-lattice $A$ ($B$). See figure \ref{fig:lat} for the sub-lattice labeling convention used here. Similarly, the Hamiltonian for the arm-chair ribbon of width $W$ is given by:
\begin{multline}
	H = \sum_i\sum_{j\in\text{odd}}^{W} t\left( c^\dagger_{i+1, j, A} c_{i, j, B} \right) + \text{h.c} \\ + \sum_i\sum_{j\in\text{even}}^{W} t\left( c^\dagger_{i, j, A} c_{i, j, B} \right) + \text{h.c} \\ + \sum_i\sum_{j=1}^{W-1} t\left( c^\dagger_{i, j+1, B} c_{i, j, A} + c^\dagger_{i, j+1, A} c_{i, j, B} \right) + \text{h.c}.
	\label{eqn:ham_ac}
\end{multline}
In this case, the extra index $j$ labels the vertical position of the site within the unit cell of the arm-chair graphene ribbon. The convention used for labeling the sites within a unit cell is shown in figure \ref{fig:lat}. For both the 1D chain and the arm-chair ribbon, since there are no terms breaking the degeneracy between the spins, only one spin species is considered for simplicity.

\begin{figure}
	\centering
	\includegraphics[width=\linewidth]{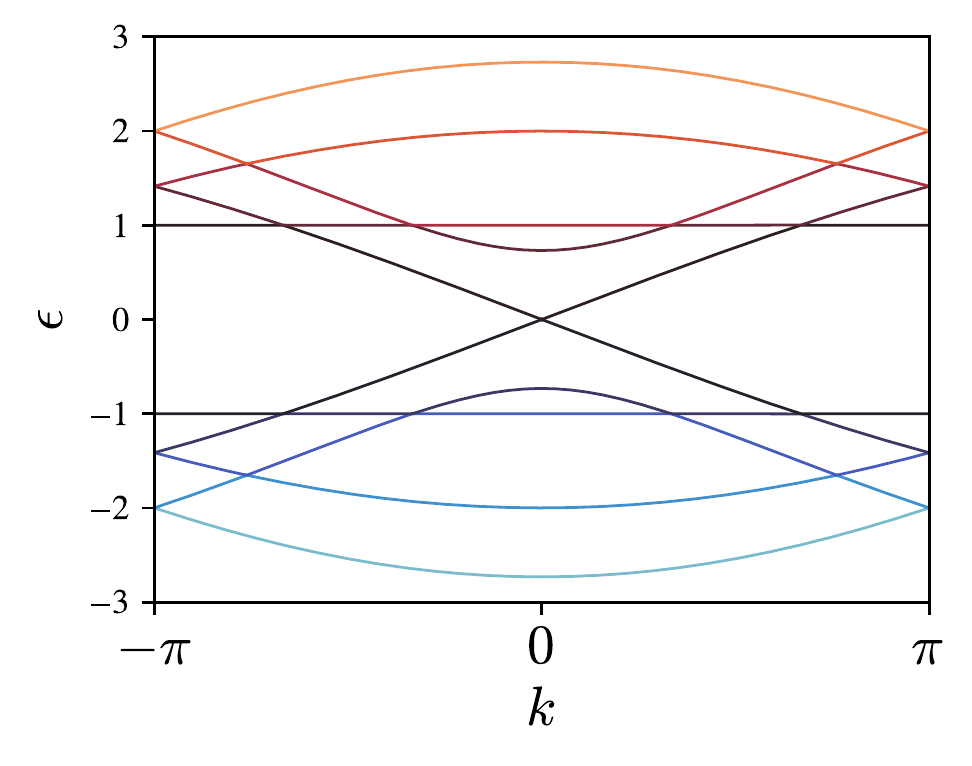}
	\caption{The spectrum of a metallic arm-chair ribbon with $W=5$ is shown, where $k$ is the momentum along the ribbon in units of the inverse lattice constant of the arm-chair ribbon. There is a linear band crossing with only one left/right moving mode close to zero energy.}
	\label{fig:ac_spectrum}
\end{figure}

The 1D chain consists of a pair of left and right moving modes, with a linear dispersion close to zero energy. The spectrum of the arm-chair ribbon can be thought of as a 1D projection of the dispersion of 2D graphene at quantized values of momentum along the transverse direction of the arm-chair ribbon \cite{WakabayashiGrapheneStates}. The dispersion generally has a gap that is inversely proportional to the width $W$. However, for widths $W = 3 n -1$, where $n$ is an integer, the dispersion contains the Dirac point of graphene and is metallic with a linear band crossing at zero energy. The spectrum of a metallic arm-chair ribbon with $W=5$ is shown in figure \ref{fig:ac_spectrum}.

Adding a random component to the hopping introduces disorder without violating the chiral symmetry that is integral to the problem, as will be shown.  In the rest of this work, a disorder strength of $V$ denotes that a random number chosen from the uniform distribution spanning $[-V, V]$ is added to every non-vanishing hopping matrix element in the Hamiltonian.

One can now examine the symmetries of the general disordered systems in more detail.  These are the three generic symmetries: time-reversal ($T$), particle-hole ($C$), and chiral ($S$). These symmetries are most easily understood by looking at the so-called first-quantized Hamiltonian $\mathcal{H}$. $\mathcal{H}$ can be thought of as a matrix whose elements $\mathcal{H}_{\alpha,\beta}$ are the co-efficients of $c^{\dagger}_\alpha c_\beta$ in the full second-quantized Hamiltonian $H$ of equation \ref{eqn:ham_1d} or \ref{eqn:ham_ac}, where $\alpha$ and $\beta$ label any site in the system. Time-reversal symmetry requires that:
\begin{equation}
	U_T^\dagger \mathcal{H}^* U_T = \mathcal{H},
\end{equation}
where $U_T$ is a unitary matrix. For our purposes, it suffices to have any anti-unitary symmetry with these properties. Thus we may consider a TRS operator that acts trivially on spin, and include in our model only a single spin species.   The corresponding time-reversal operator acts trivially on the fermion creation and annihilation operators, i.e. $U_T = \mathds{I}$. Therefore, time-reversal symmetry is present provided that all matrix elements of $\mathcal{H}$ are real.  In this case, we necessarily find $T^2=+1$.

One of the most important symmetries present in these systems is chiral symmetry. Chiral symmetry requires that:
\begin{equation}
	U_S^\dagger \mathcal{H} U_S = -\mathcal{H},
\end{equation}
where $U_S$ is a unitary matrix. In this case, the chiral symmetry is given by:
\begin{equation}
	(U_S)_{\alpha,\beta} = 
	\begin{cases}
		0, & \alpha \neq \beta \\
		1, & \alpha=\beta \text{ belongs to sub-lattice A} \\
		-1, & \alpha=\beta \text{ belongs to sub-lattice B}
	\end{cases}
\end{equation}
It follows that chiral symmetry is present when the Hamiltonian only contains hopping terms between different sub-lattices, which is always the case here. Finally, combining time-reversal and chiral symmetry yields particle-hole symmetry $C$, that squares to $+1$. This puts the systems in symmetry class BDI.

The package KWANT for python \cite{KWANT} is used to compute the tranport properties numerically. This package enables the creation of finite size tight binding systems with disorder. One can then attach semi-infinite leads of the corresponding clean system and then compute the S-matrix using methods available in KWANT. From the S-matrix, one can obtain the transmission eigenvalue(s) and therefore $x$ by using equation \ref{eqn:x_definition}. Note that this constrains $x \geq 0$, so the distribution class BDI \ref{eqn:bdi_prob} needs to be folded in order to perform a comparison with numerics. This enables the study of transport with a variety of sizes, energies, and disorder strengths. Over 20000 disorder realizations are computed for each value of these parameters. For more information on the inner workings of the package, such as how quantities are computed, readers are referred to the aforementioned reference and the software documentation.

\section{Results}  \label{Sec:Results}

In this section, 
the main results of our work are presented. First, the existence of a two-parameter scaling of transport distributions at finite energy near the zero energy critical point of class BDI is demonstrated in \ref{subsec:scaling}. This is done by showing that the numerically obtained  disorder averaged transport quantities for the 1D chain \ref{eqn:ham_1d} exhibit scaling collapse. The resulting two-parameter scaling interpolates between the transport statistics of the chiral distribution for class BDI \ref{eqn:bdi_prob} at zero energy to that of class AI \ref{eqn:ai_prob} at large energies. Next, the behavior of the two-parameter probability distribution $P(x; s, r)$ is studied in \ref{subsec:prob_dsbn}. It is shown that there are three distinct regimes of transport, as outlined in figure \ref{fig:phase_portrait}. It is also shown that the overall shape of the distribution is controlled by the ratio $s/\abs{\ln(r)}^2$, which is related to the ratio of the average localization length and the system size. Finally, in \ref{subsec:univ}, the apparent universality of the two-parameter scaling is discussed, along with a brief comparison of the results of this work to that of ref. \cite{RyuCrossover}.

\subsection{Establishing two-parameter scaling} \label{subsec:scaling}

To understand the cross-over of transport from class BDI to AI, we begin by studying the 1D chain. As seen in the background section \ref{Sec:Background}, the FP equations imply that the transport distribution $P(x)$ in a given symmetry class, such as BDI at $\epsilon=0$ or AI, is completely determined by a single parameter $s = \frac{L}{l}$. Conversely, one can obtain $s$ from from a single disorder averaged quantity, usually $-\expval{\ln(g)}$. In contrast, the transport statistics of our models in class BDI away from zero energy obey a 2-parameter scaling, with the second parameter necessarily depending on the energy. This is because one expects the transport distributions to cross-over from that of class BDI (eqn. \ref{eqn:bdi_prob}) at zero-energy to class AI (eqn. \ref{eqn:ai_prob}) at large energies. Below, numerical evidence is presented showing that transport is a function of the two parameters $s$ and $r = \epsilon \tau$. More concisely:
\begin{equation}
	P(x) \equiv P(x; s, r).
\end{equation}
Note that there are three independent parameters: the length $L$, energy $\epsilon$, and disorder strength $V$, but in practice, only two dimensionless parameters control the transport distributions.

\begin{figure}
	\centering
	\includegraphics[width=\linewidth]{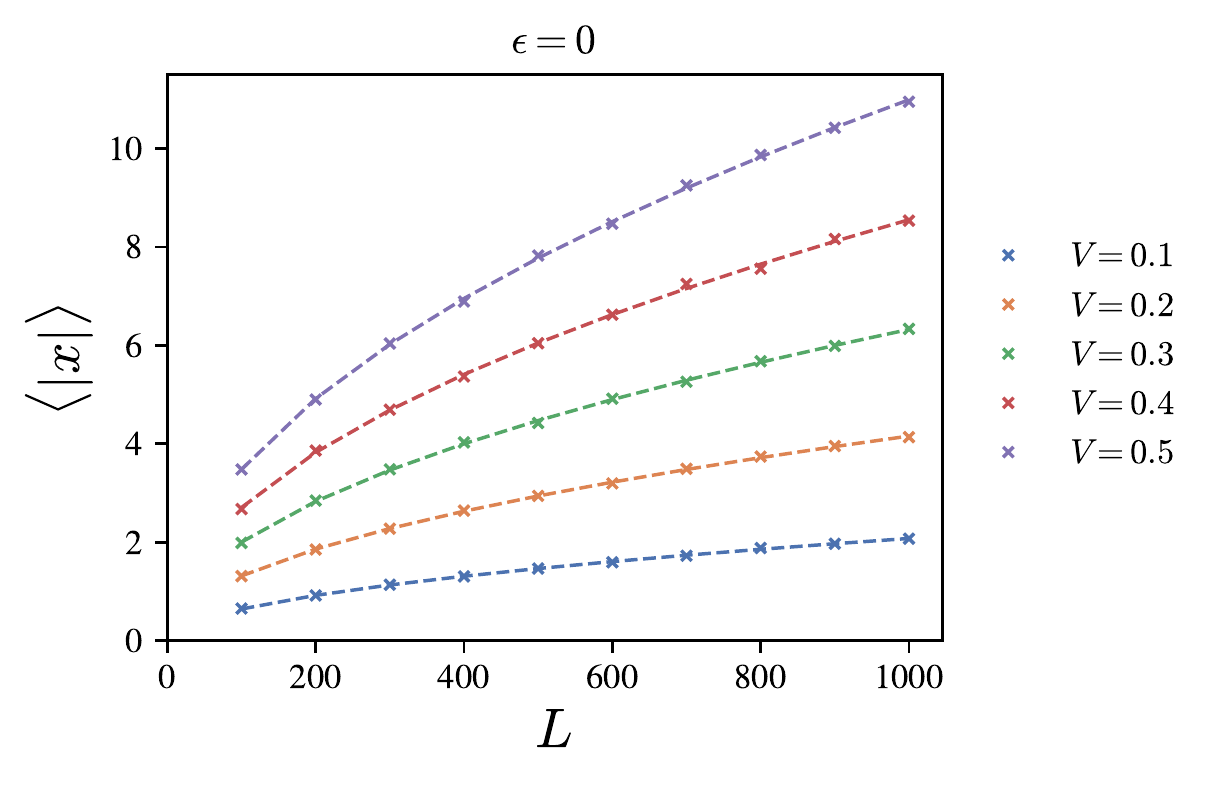}
	\caption{The plot shows $\expval{\abs{x}}$ vs $L$ for several disorder strengths at zero energy in the 1D chain. The dashed lines are fits to the prediction from equation \ref{eqn:bdi_prob}, which gives $\expval{\abs{x}} = \sqrt{\frac{2 s}{\pi}}$. The fits show good agreement and $l$ is extracted from this.}
	\label{fig:zero_energy_avgx}
\end{figure}

The first step in showing the two parameter scaling is to obtain the mean free path $l$ from the quantity $\expval{\abs{x}}$, where $\expval{\dots}$ denotes disorder averaging, at zero energy, for several disorder strengths. This is shown in figure \ref{fig:zero_energy_avgx}, with the fits showing good agreement with the prediction obtained using equation \ref{eqn:bdi_prob}. From this, one can also obtain $\tau = \frac{v_F}{l}$. Over the energy ranges studied here ($\epsilon \in [10^{-6}, 10^{-2}]$), it is safe to assume that $l$ and $\tau$ are nearly independent of energy. This can be justified from a simple Fermi's golden rule calculation for a single weak impurity. In this approximation, the scattering rate is given by:
\begin{equation}
	\tau^{-1}(\epsilon) = 2\pi \, \rho_0(\epsilon) \, \abs{\mel{\psi_L}{H_{imp}}{\psi_R}}^2,
\end{equation}
where $H_{imp}$ is the Hamiltonian of an impurity of strength $V$ located on a single bond, $\psi_L (\psi_R)$ is the wavefunction of the left(right) moving mode at a given energy, and $\rho_0(\epsilon)$ is the density of states of the clean system. Evaluating the matrix element gives:
\begin{equation}
	\tau^{-1}(\epsilon) = \frac{2 V^2}{L v_F}.
\end{equation}
where $v_F$ is the group velocity. Thus the energy dependence of $\tau$ and $l$ essentially comes from the Fermi velocity, which is well approximated by a constant for the energies used here.

\begin{figure}
	\centering
	\includegraphics[width=\linewidth]{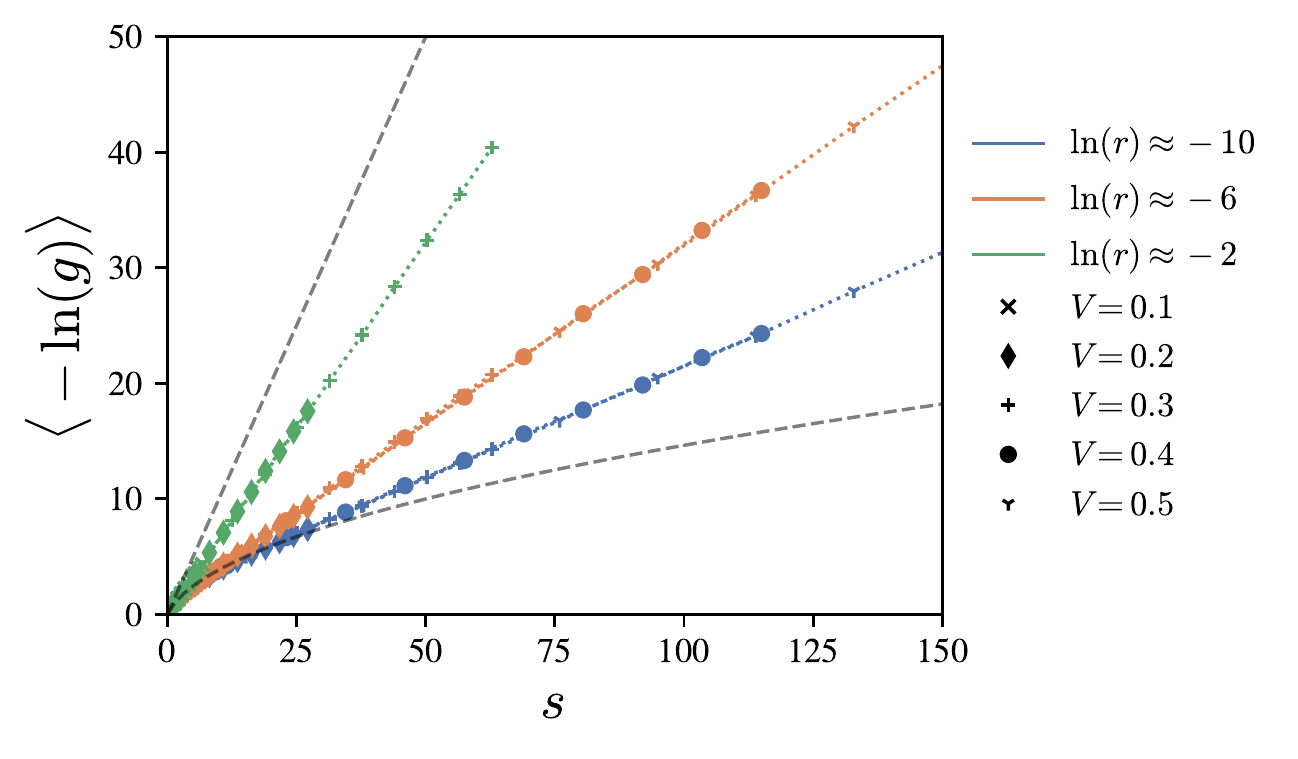}
	\caption{Two parameter data collapse of $\expval{-\ln(g)}$ for the 1D chain with several disorder strengths. The scaling parameter $s$ is given by the $x$-axis, whereas the different colors span the second scaling parameter $r$. The markers correspond to different disorder strengths and the dotted lines simply join neighboring points to clearly show collapse. The upper (lower) dashed lines are the predictions from the FP equation for class AI (BDI at zero energy) that should be applicable for large (zero) $r$. The lengths of the systems range from 10 - 1000 unit cells and the energies span 10$^{-6}$ - 10$^{-2}$.}
	\label{fig:lng_collapse}
\end{figure}

\begin{figure}
	\centering
	\includegraphics[width=\linewidth]{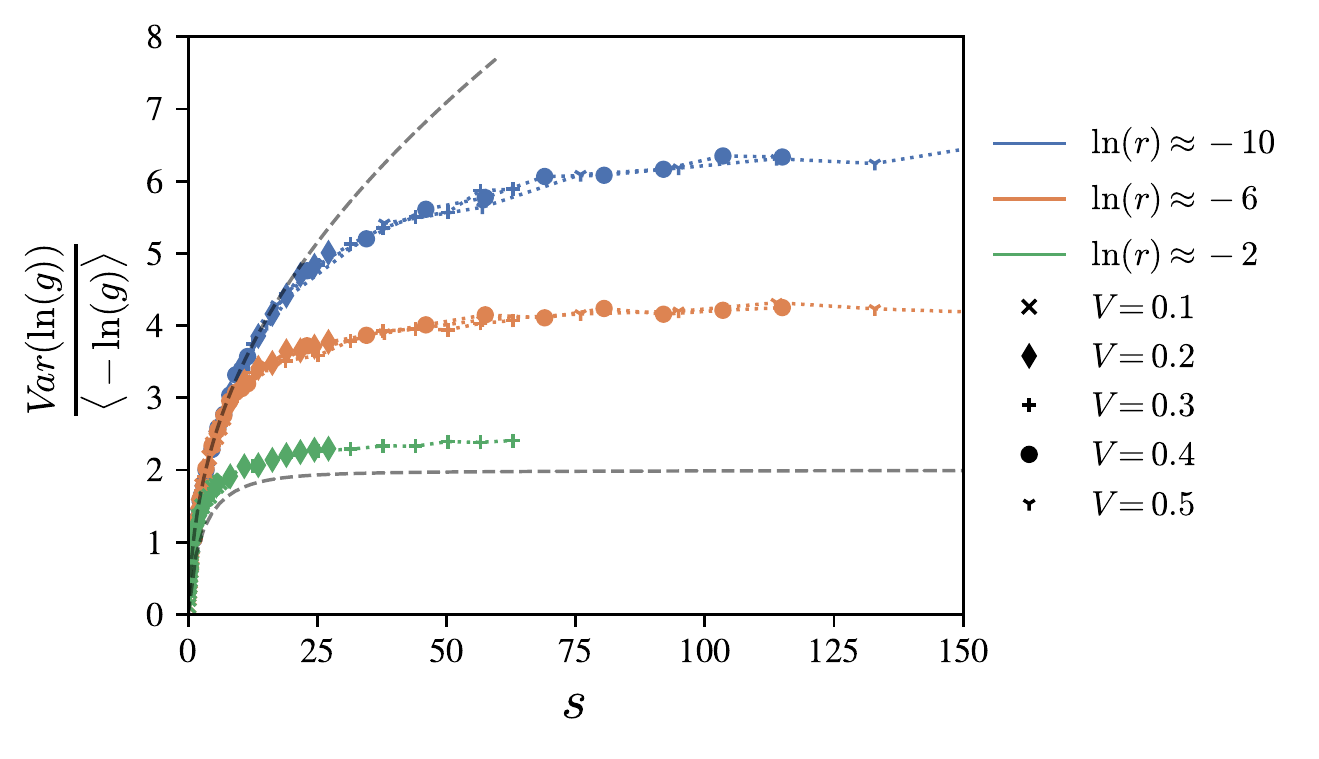}
	\caption{Similar to figure \ref{fig:lng_collapse}, but for the quantity $\frac{\mathrm{Var}(\ln(g)}{\expval{-\ln(g)}}$. Once again, there is good agreement with the two-parameter scaling hypothesis. Here the upper (lower) dashed lines correspond to class BDI (AI). Observe that the curves seem to saturate at increasingly large values as the energy is lowered.}
	\label{fig:varg_collapse}
\end{figure}

One can now show numerically that $P(x)$ is completely specified by $s$ and $r$. A good way to visualize two parameter scaling across a range of parameters is by observing data-collapse for a few key quantities when plotted as a function of $s, r$. Figures \ref{fig:lng_collapse} and \ref{fig:varg_collapse} show such data collapse for the quantities $\expval{-\ln(g)}$ and $\frac{\mathrm{Var}(\ln(g)}{\expval{-\ln(g)}}$, demonstrating excellent agreement with the postulated two-parameter scaling. The limiting values of $\expval{-\ln(g)}$ at large and zero $\epsilon \tau$ (Fig. \ref{fig:lng_collapse}, black dashed lines) is given in equations \ref{eqn:lng_ai} and \ref{eqn:lng_bdi}, respectively.

The behavior of $\frac{\mathrm{Var}(\ln(g)}{\expval{-\ln(g)}}$ also highlights the growing contribution from rare regions at lower energies, which lead to large conductance fluctuations. As mentioned in section \ref{Sec:Background}, these rare-regions are formed due to the formation of domains in topologically distinct phases with bound states on domain walls. For class AI it is expected that this quantity asymptotically approaches 2 for large $s$ based on equation \ref{eqn:ai_prob}. In contrast, for class BDI, where rare-region effects dominate\cite{BalentsFisher,MotrunichDamleHuse}, for $s \gg 1$ Eq. \ref{eqn:bdi_prob} gives:
\begin{equation}
	\frac{\mathrm{Var}(\ln(g)}{\expval{-\ln(g)}} \approx \left(1 -\frac{2}{\pi}\right) \sqrt{2\pi s,}
\end{equation}
which never saturates, indicating that the system is not self-averaging.  As shown in figure \ref{fig:varg_collapse}, for any non-zero $r$ this quantity does saturate, but at a value that increases  with $- \ln r$. This, combined with the observation that $\expval{-\ln(g)}$ behaves linearly with $s$ for sufficiently large $s$ and all observed values of $r$, indicates that $\ln(g)$ is self-averaging for sufficiently large systems at any finite energy.

\begin{figure}
	\centering
	\includegraphics[width=\linewidth]{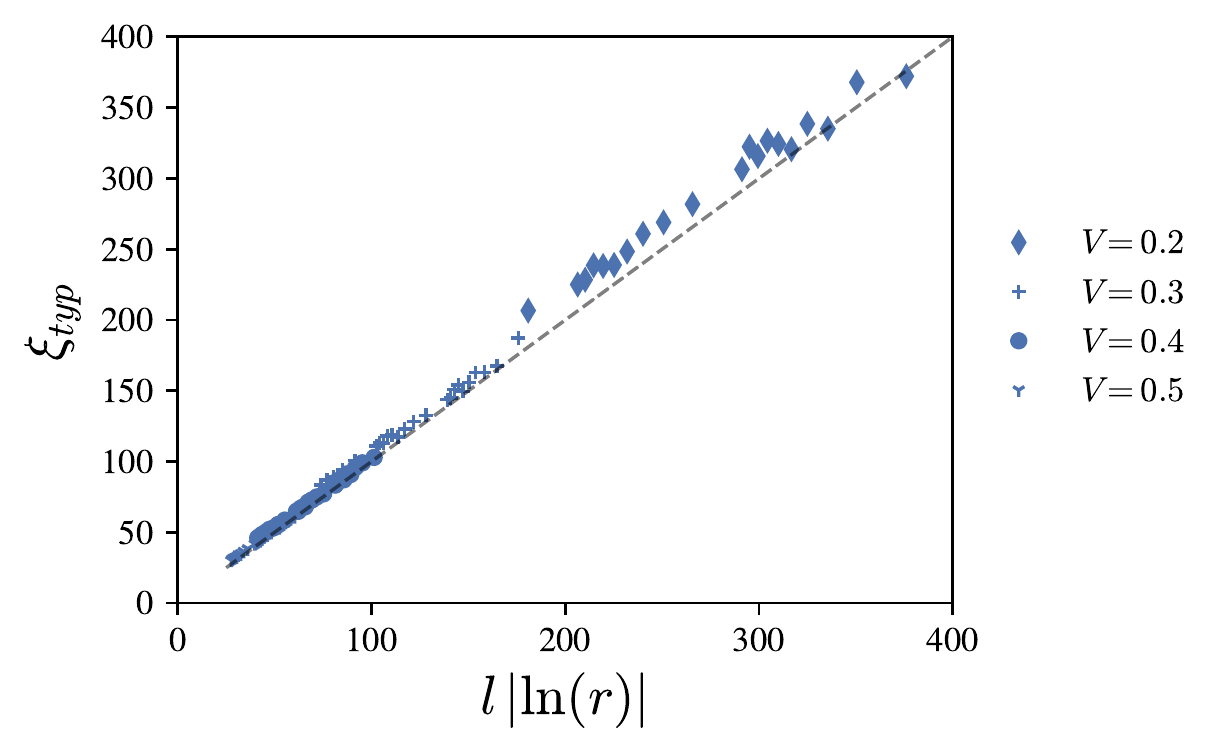}
	\caption{The form of the divergence of the typical localization length with energy is shown. The dashed line shows the expected scaling near the 0-energy critical point, $\xi_{typ} = l |\ln r|$. $\xi_{typ}$ is obtained numerically by fitting to equation \ref{eqn:xi} for sufficiently large system size $L$. $r < 10^{-2}$ for all points in the figure. The data is noisier when the localization length becomes comparable to the largest system size (1000 unit cells).}
	\label{fig:xi}
\end{figure}

Still, the increasing saturation value of $\frac{\mathrm{Var}(\ln(g)}{\expval{-\ln(g)}} $ at smaller $r$ indicates the relative broadening of the distribution as the contribution of rare regions to transport becomes increasingly important.  
Such behavior was predicted by Refs. \cite{BalentsFisher, MotrunichDamleHuse}, who found a growing discrepancy between \emph{typical} and \emph{average} quantities at lower energies.   For instance, it was predicted that the typical localization length $\xi_{typ}$ goes as $\abs{\ln(\epsilon)}$, whereas the average localization length $\xi_{avg}$ goes as $\ln^2(\epsilon)$, when $r \ll 1$ \cite{McKenzieQPT, BalentsFisher}. 
Figure \ref{fig:xi} indicates that numerically, the typical localization length goes as:
\begin{equation}
	\xi_{typ} = l \abs{\ln(r)}
	\label{eqn:xi_typ}
\end{equation}
when $r \ll 1$, in line with the results of \cite{McKenzieQPT}.

\subsection{Characterizing the distribution $P(x; s, r)$} \label{subsec:prob_dsbn}

Having established the two-parameter scaling of transport statistics, one can investigate the form of $P(x)$ for different values of $(s, r)$.   Numerically, the distribution $P(x)$ is extracted from the distribution of the conductance $g = 1/\cosh^2(x)$.  For a given $g$, this in general yields two solutions, one at $x>0$ and one at $x<0$.  Here we always take the positive solution, such that the  numerically obtained distribution $P(x)$ is defined only for $x>0$.  Since in the chiral class the FP equation allows any $- \infty < x < \infty$, our numerical distributions are thus folded relative to the true distributions, with any probability formally associated with $x<0$ contributing to $P(-x)$.

\begin{figure}
	\centering
	\includegraphics[width=\linewidth]{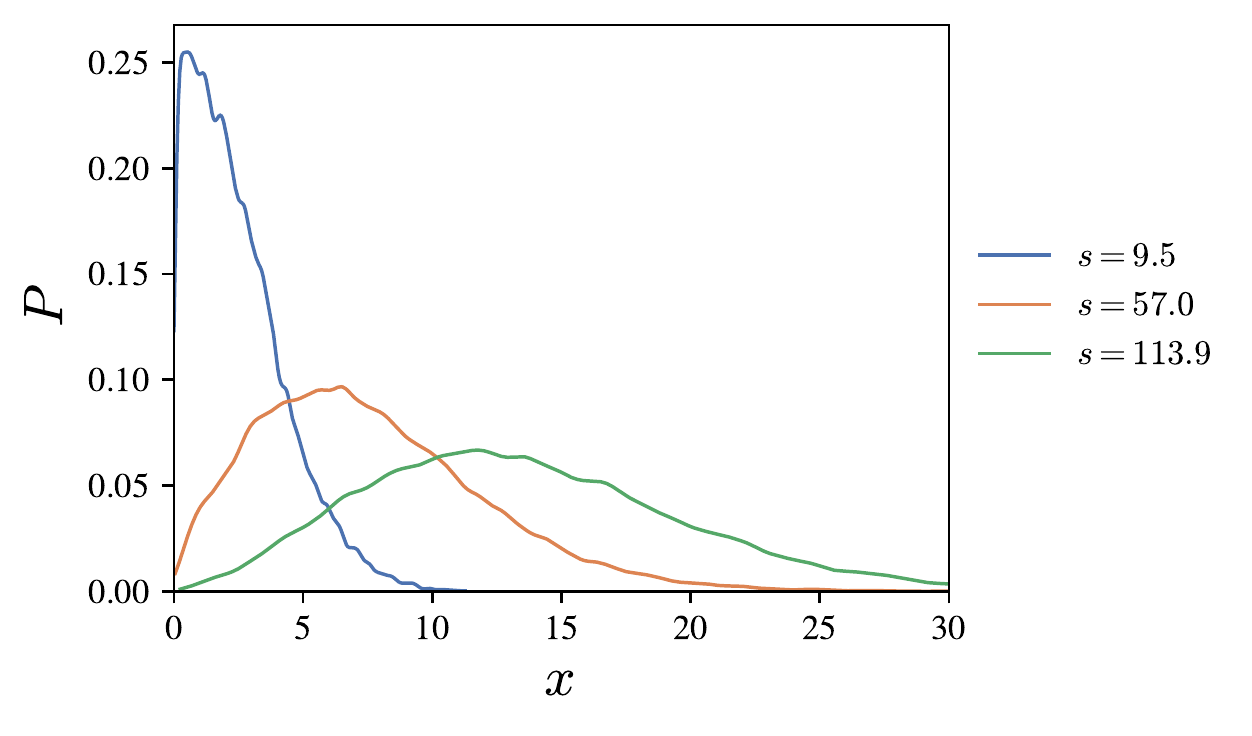}
	\caption{The evolution of the probability distribution $P(x)$ with $s$ at fixed $\ln(r) \approx -9.2$ is shown. $V = 0.5$ for all lines.}
	\label{fig:prob_dsbn_vs_s}
\end{figure}

Figure \ref{fig:phase_portrait} summarizes the evolution of $P(x)$ with $(s,r)$.  Three different transport regimes can be identified, depending on the relative magnitudes of $r$ and $s$: (1) the localized regime, which occurs at sufficiently large $s$ for any non-zero $r $, (2) the chiral regime, which occurs at sufficiently small $s$ for all values of $r$ studied here, and (3) the cross-over regime, which connects the two.  Each regime corresponds to a qualitatively different form of the distribution $P(x)$, as shown in Figure \ref{fig:prob_dsbn_vs_s}.
We now discuss each of them in detail.

\subsubsection{Localized regime}

The localized regime, in the upper right of figure \ref{fig:phase_portrait}, describes parameter values where $\expval{-\ln(g)}$ grows linearly with $s$, and $\frac{\mathrm{Var}(\ln(g))}{\expval{-\ln(g)}}$ is approximately independent of $s$.  For $s \gg 1$, where this regime occurs at small $r$, this suggests that $P(x)$ becomes a Gaussian distribution of the form:
\begin{equation}
	P(x) \approx \frac{1}{\sqrt{2 \pi \sigma^2}} \exp(-\frac{(x - \mu)^2}{2 \sigma^2}).
	\label{eqn:gaussian_dsbn}
\end{equation}
Such a distribution agrees qualitatively with the predictions for class AI (see Eq. (\ref{eqn:ai_prob})) at large $s$, but with a different mean and variance. In this regime, the distribution has negligible weight near $x=0$, as implied by the form of the normalization in equation \ref{eqn:gaussian_dsbn}. This requires that $\mu \gg \sigma$ for self-consistency.

\begin{figure}
	\centering
	\includegraphics[width=\linewidth]{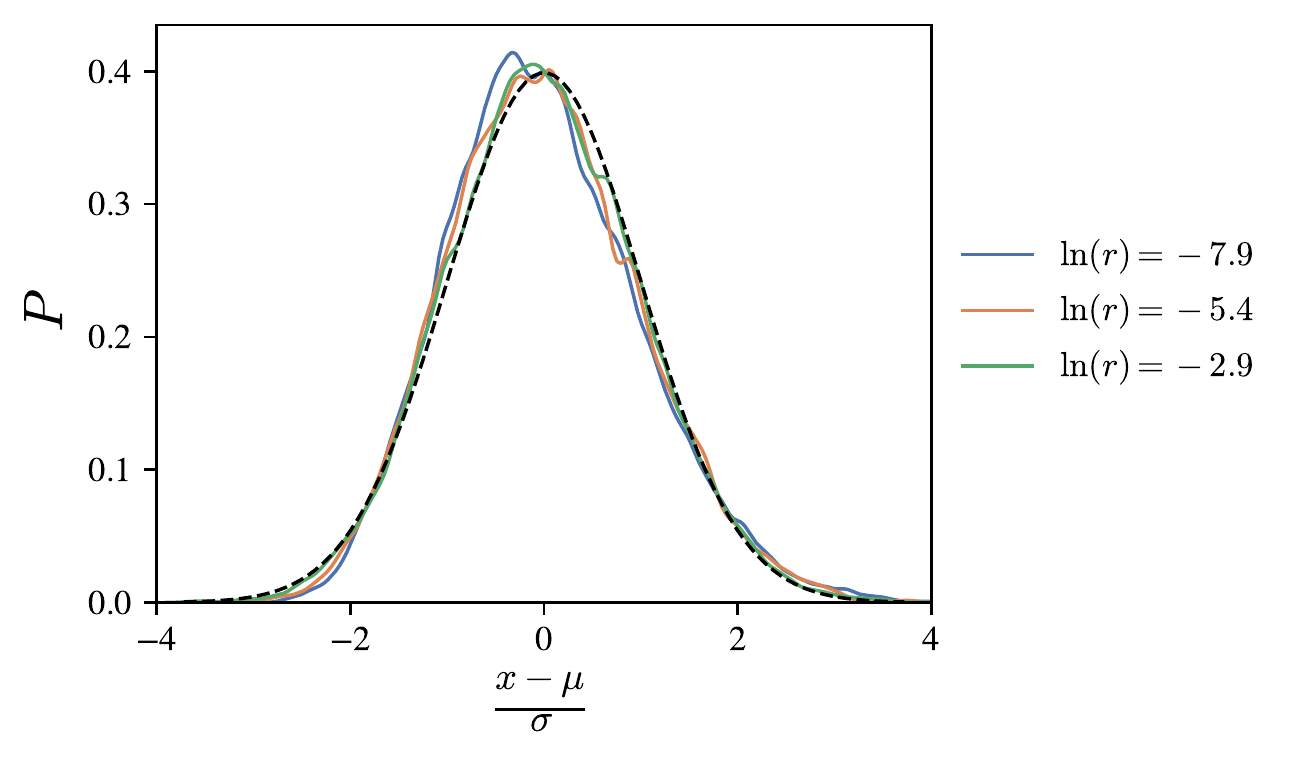}
	\caption{The probability distribution for $x$ after subtracting the mean and scaling by the standard deviation for $s \gg |\ln r|^2$. The black dashed line is the standard normal distribution. $V = 0.5$ and $s \approx 190$ are used for all lines.}
	\label{fig:prob_dsbn_norm}
\end{figure}

\begin{figure}
	\centering
	\includegraphics[width=\linewidth]{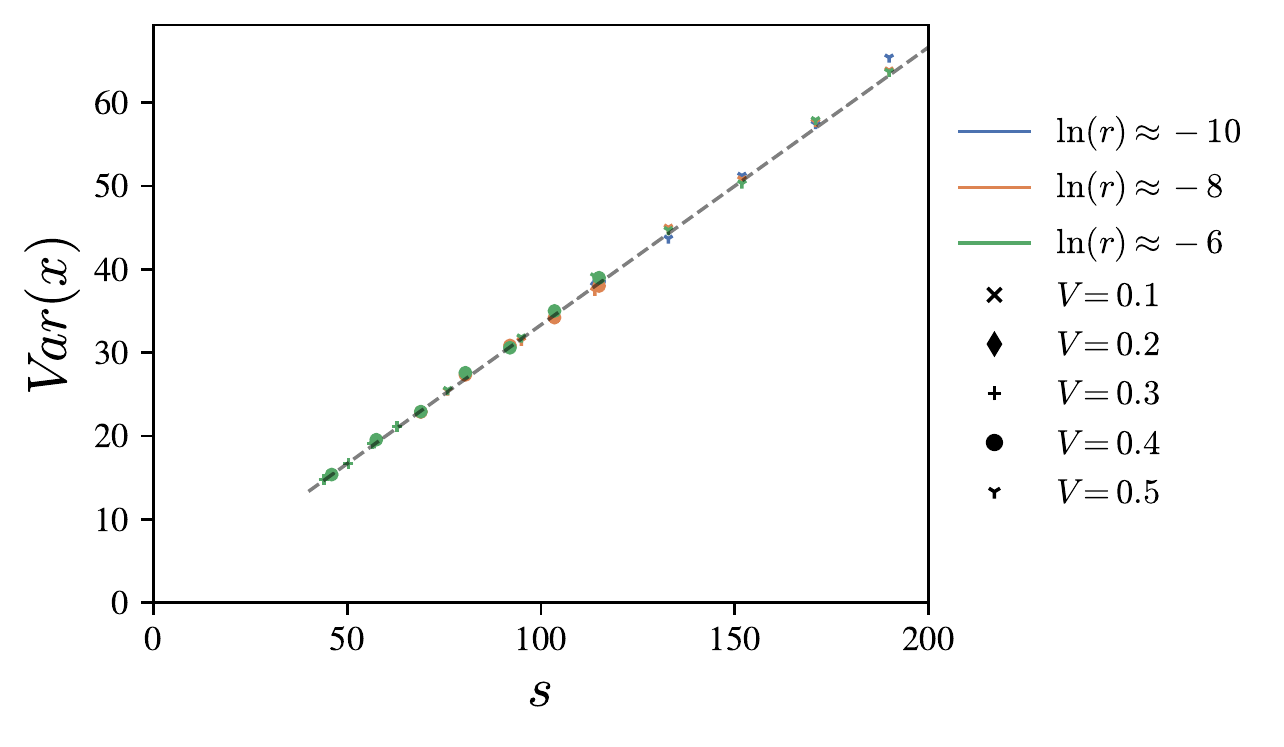}
	\caption{The variance of $x$ is shown as a function of s when $s \gg 1$. The dashed line shows $\frac{s}{3}$ as a reference.}
	\label{fig:varx}
\end{figure}

Figure \ref{fig:prob_dsbn_norm} compares the numerical distributions obtained at $s = 190$ to the Gaussian form (\ref{eqn:gaussian_dsbn}).  The plot shows good agreement, indicating that $P(x)$ is well-approximated by a Gaussian, though the distribution is exactly Gaussian only in the limit $s \to \infty$.  (This is true even for Class AI.)  The parameters $\mu, \sigma$ can be extracted from the numerics as follows. Using equation \ref{eqn:gaussian_dsbn} one finds:
\begin{equation}
	\expval{-\ln(g)} \approx 2 \mu
\end{equation}
Combining this with the result of equation \ref{eqn:xi_typ}, one finds that for $s \gg 1$ and $r \ll 1$:
\begin{equation}
	\mu = \frac{s}{\abs{\ln(r)}} + \mathcal{O}(\ln(r)) .
	\label{eqn:mu}
\end{equation}
The term of $\mathcal{O}(\ln(r))$ can be ignored when $s$ is sufficiently large. 
The variance can be extracted directly from the numerical results, as shown in \ref{fig:varx}.  The corresponding best fit line yields: 
\begin{equation}
	\sigma^2 \approx \frac{s}{3}.
	\label{eqn:sigma}
\end{equation}

The scale at which the system enters the localized regime can be understood 
based on the observation that our distribution will be Gaussian only when the condition $\mu \gg \sigma$ is satisfied. From Eqs. (\ref{eqn:mu}) and (\ref{eqn:sigma}), this implies that $s \gg \ln^2(r)$, for small $r$.  Therefore the length scale to enter the localized regime for small $r$ is given by:
\begin{equation} \label{Eq:GaussCross}
	L \gg \frac{\xi_{typ}^2}{l} \sim \xi_{avg}.
\end{equation}
This length scale was also identified as relevant to the onset of the cross-over regime in multi-mode systems by Ref. \cite{GruzbergSuperuniversality}.

\subsubsection{Chiral regime}

\begin{figure}
	\centering
	\includegraphics[width=\linewidth]{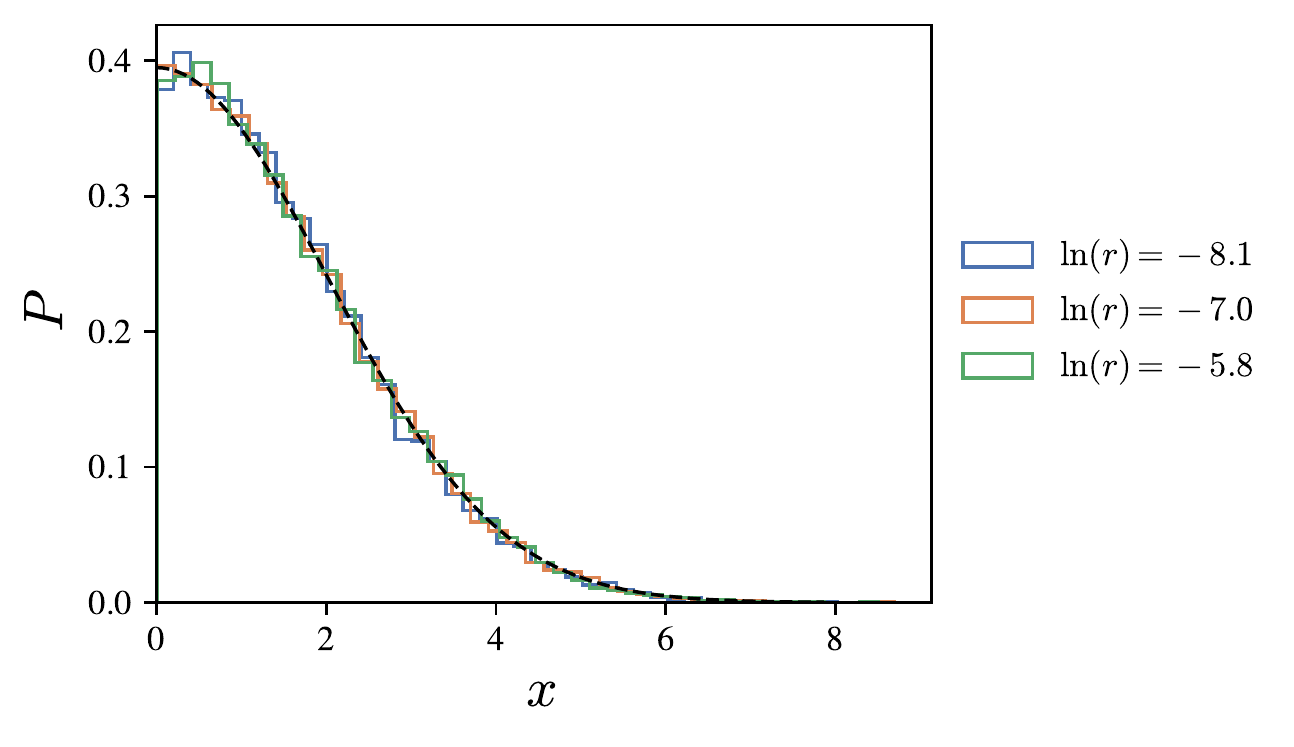}
	\caption{The probability distribution $P(x)$ for a few different values of $r$ in the small $s$ limit where the distribution resembles that of class BDI. $s \approx 4$ and $V=0.1$ are used. The black dashed line shows equation \ref{eqn:bdi_prob} with this value of $s$.}
	\label{fig:prob_dsbn_chiral}
\end{figure}

The far left  of figure \ref{fig:phase_portrait} shows the Chiral regime, which occurs for   sufficiently small $s$ and $r \lesssim 1$.  Here the distribution $P(x)$ is  essentially independent of $r$, and indistinguishable from that of equation \ref{eqn:bdi_prob}, which describes transport in class BDI at zero energy. This is shown in figure \ref{fig:prob_dsbn_chiral}.   Appendix \ref{Sec:StatTests} discusses a $p$-value test confirming the proximity of $P(x)$ to the chiral distribution in this region. Note that the numerics are restricted to $x \geq 0$, so the distribution of equation \ref{eqn:bdi_prob} needs to be folded in order to perform this comparison. As seen in figure \ref{fig:phase_portrait}, the cut-off length scale below which the transmission statistics are well-described by Eq. (\ref{eqn:bdi_prob}) decreases as $r$ increases.  This is consistent with the observation that the threshold lengthscale for the system to be self-averaging also decreases with $r$.  We return to this point presently.

\subsubsection{Crossover regime}

\begin{figure}
	\centering
	\includegraphics[width=\linewidth]{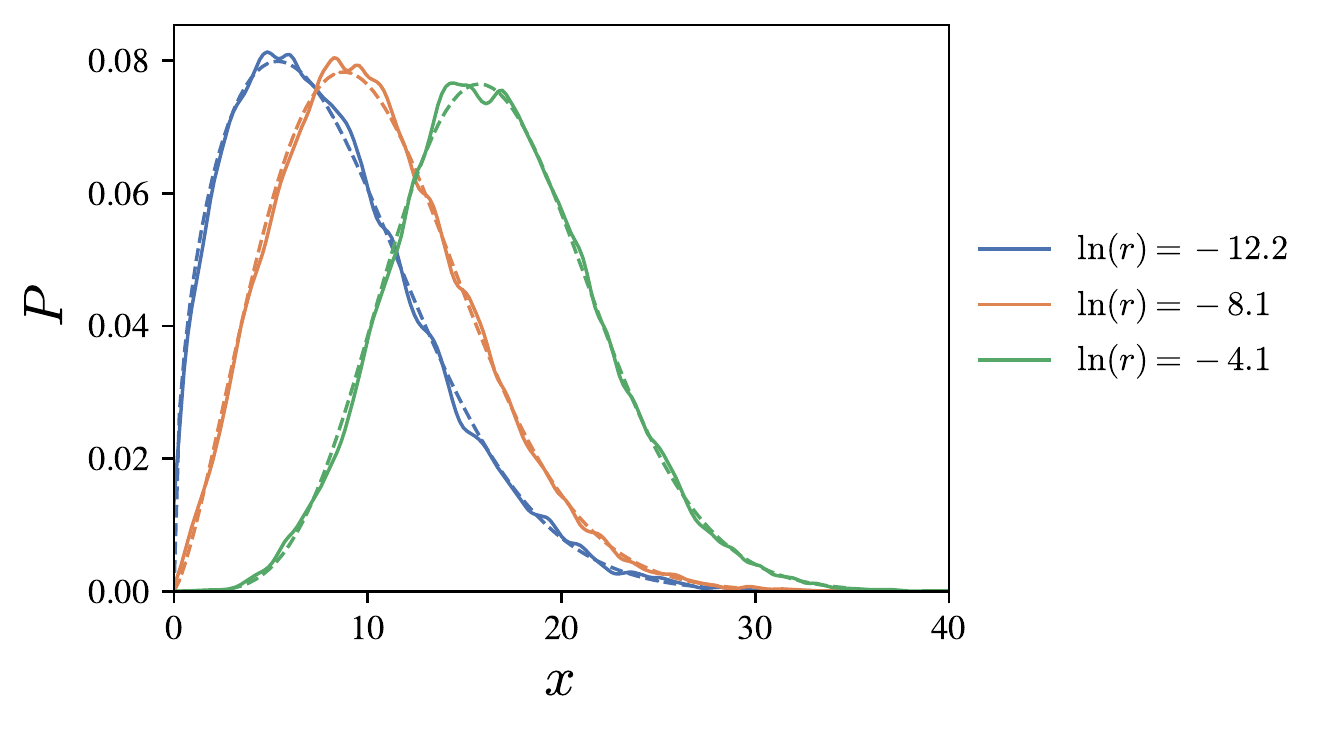}
	\caption{$P(x)$ at several values of $r$, with $s = 76$ and $V=0.5$, showing good agreement with equation \ref{eqn:gamma_dsbn}. The solid lines are from numerics whereas the dashed lines are fits to equation \ref{eqn:gamma_dsbn}.}
	\label{fig:prob_dsbn_gamma}
\end{figure}

In the cross-over region between the chiral and localized regimes, the numerical distribution $P(x)$ is well approximated by a distribution of the form:
\begin{equation}
	P(x) = \frac{2}{\gamma \, \Gamma(\delta/2)} \left( \frac{x}{\gamma}\right)^{\delta-1} \exp\left[- \left( \frac{x}{\gamma} \right)^2 \right], \ x\geq 0
	\label{eqn:gamma_dsbn}
\end{equation}
where the parameters $\delta, \gamma$ are functions of $s, r$. This is a special case of the generalized gamma distribution known as the Nakagami distribution. Figure \ref{fig:prob_dsbn_gamma} shows the strong agreement between the ansatz \ref{eqn:gamma_dsbn} and the numerical results at several points in the cross-over regime.

As $\delta$ is varied, the Nakagami distribution interpolates smoothly between the folded version of distribution of class BDI at zero energy (eqn. \ref{eqn:bdi_prob}), obtained for $\delta=1, \gamma = \sqrt{2s}$, and the Gaussian distribution described by equation \ref{eqn:gaussian_dsbn}, obtained by taking $\delta \to \infty$ (See  appendix \ref{GaussApp} for a derivation). In the Gaussian limit, the parameters $\mu$ and $\sigma$ are given by
\begin{equation}
	\mu = \gamma \sqrt{\frac{\delta}{2}} \; ; \;  \sigma = \frac{\gamma}{2}.
	\label{eqn:gamma_dsbn_limit}
\end{equation}

To understand the scales at which the system passes between the chiral and cross-over regimes,
it is useful to relate the parameter $\delta$ to quantities that are relevant to transport.  To do this, note that for general $\gamma, \delta$,
\begin{equation}
	\langle x \rangle = \gamma \frac{ \Gamma \left( \frac{1 + \delta}{2} \right )}{\Gamma \left( \frac{ \delta}{2} \right )} \ , \ \ \langle x^2 \rangle = \frac{1}{2} \delta \gamma^2  \ ,
\end{equation}
which gives
\begin{equation}
	\sigma^2 = \frac{\gamma^2}{4} \left ( 2 \delta -4 \left(  \frac{ \Gamma \left( \frac{1 + \delta}{2} \right )}{\Gamma \left( \frac{ \delta}{2} \right )}  \right )^2 \right )  \approx  \frac{\gamma^2}{4} \ .
\end{equation} 
Thus the ratio 
\begin{equation} \label{Eq:Delta1}
	\frac{\langle x \rangle^2 }{ \langle x^2 \rangle - \langle x \rangle^2 } = \frac{ \Gamma \left( \frac{1 + \delta}{2} \right )^2 }{\frac{ \delta }{2} \Gamma \left( \frac{ \delta}{2} \right )^2  -  \Gamma \left( \frac{1 + \delta}{2} \right )^2 } \approx 2 \delta 
\end{equation}
is independent of $\gamma$, and is approximately equal to $2 \delta$ for $1 \leq \delta \leq \infty$, with a correction that decreases from $-0.25$ for $\delta = 1$ to $-0.5$ as $\delta \rightarrow \infty$.    In other words, $\delta$ is approximately determined by the square of the ratio of the distribution's mean and standard deviation; this correspondence becomes exact in the Gaussian limit $\delta \rightarrow \infty$, i.e. $L \gg  (\xi_{typ})^2 /l$.   
Taking 
\begin{equation}
	\langle x \rangle \approx - \frac{1}{2} \langle \log g \rangle = \frac{L}{\xi_{typ}}
\end{equation}
and using the numerically obtained value of $\sigma^2   \approx \frac{s}{3}$ gives 
\begin{equation} \label{Eq:Delta2}
	\delta \approx \frac{3}{2}  \frac{L l}{(\xi_{typ})^2}  = \frac{3}{2} \frac{s}{|\ln^2 r | }\ .
\end{equation}
where the last equality is valid for $r \ll 1$, where Eq. (\ref{eqn:xi_typ}) can be used.  
Thus the shape of the distribution is approximately controlled by the ratio of the length $L$ to the scale $(\xi_{typ})^2/l \sim \xi_{avg}$, with this relationship becoming exact for large $\delta$.  From Eq. (\ref{Eq:Delta1}), this ratio measures the extent to which the system is self-averaging.

Setting  $\delta\approx 1$ in Eq. (\ref{Eq:Delta2}) suggests that the length scales below which the Chiral regime occurs is expected to be:
\begin{equation}
	L \lesssim  \frac{2}{3}  \frac{\xi_{typ}^2}{l}.
\end{equation}

It is interesting to compare these predictions to the  numerical phase diagram shown in figure \ref{fig:phase_portrait}.  The color map in this phase diagram corresponds to the skewness of $x$, denoted by $\tilde{\mu}_3(x) = \expval{(x - \mu)^3/\sigma^3 }$, where $\mu$ and $\sigma^2$ are the mean and variance of the distribution $P(x)$ respectively.  When the Chiral distrbution of equation \ref{eqn:bdi_prob} is folded to $x > 0$, $\tilde{\mu}_3(x)$ is almost $1$. In addition, $\tilde{\mu}_3(x)$ falls to zero for the Gaussian distribution of equation \ref{eqn:gaussian_dsbn}.   Based on this, approximate boundaries for the regimes are obtained in figure \ref{fig:phase_portrait}. The blue and red dashed lines separating the regimes in this figure are given by $L = 0.28 \frac{\xi_{typ}^2}{l}$ and $L = 2 \frac{\xi_{typ}^2}{l}$. This is in agreement with the expectations mentioned above.

\subsection{Universality of two-parameter scaling and comparison to ref. \cite{RyuCrossover}} \label{subsec:univ}

\begin{figure}
	\centering
	\includegraphics[width=\linewidth]{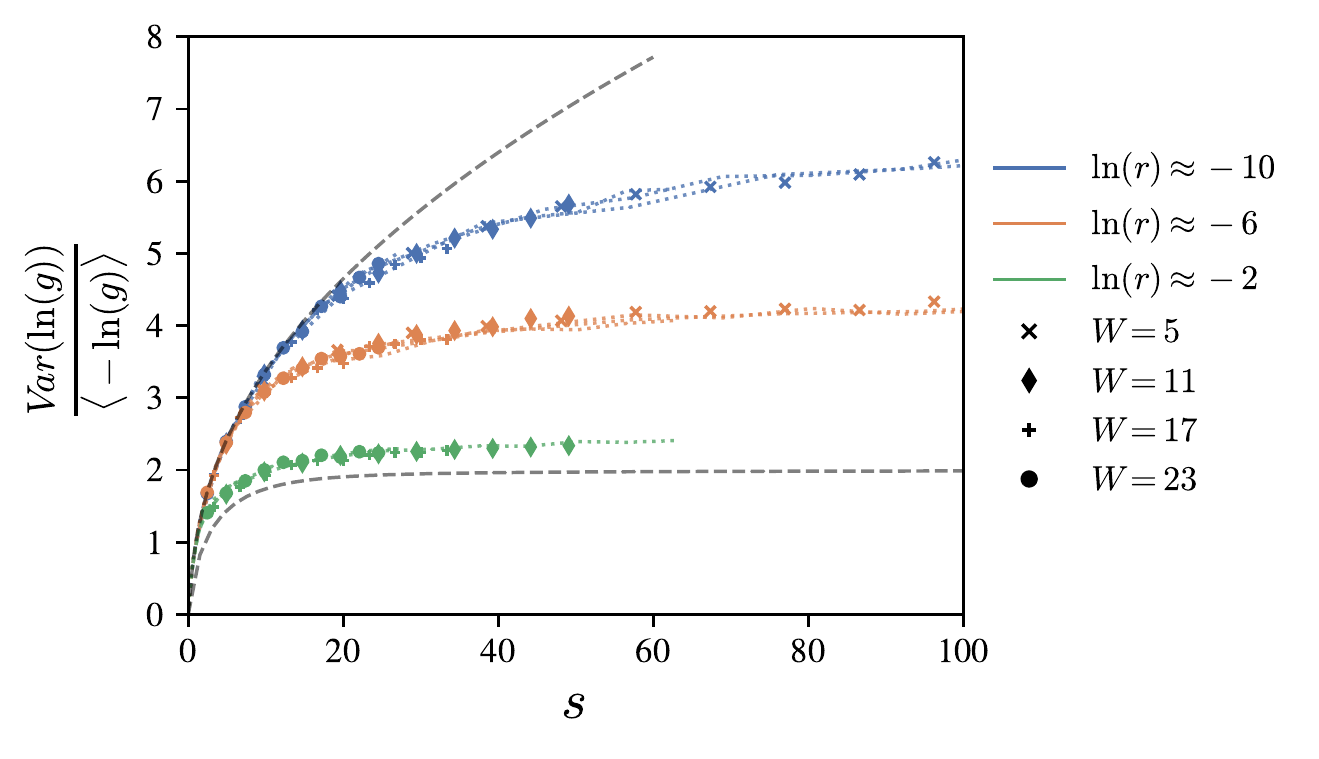}
	\caption{Two parameter data collapse similar to figure \ref{fig:varg_collapse}, but for metallic arm-chair ribbons of different widths with $V=0.5$. The dotted lines here are from the 1D chain at the corresponding value of $\epsilon\tau$. The ribbons have lengths of 100 - 1000 unit cells and energies from 10$^{-6}$ - 10$^{-2}$.}
	\label{fig:varg_collapse_ac}
\end{figure}

If the two-parameter scaling of $P(x)$ describes a cross-over away from criticality, one might expect it to be a universal property of the critical point, rather than a feature of the specific model.  A modest test of such universality is whether it holds for other 1D systems with a single linearly dispersing mode close to zero energy, such as metallic arm-chair graphene nanoribbons at sufficiently low energies. Figure \ref{fig:varg_collapse_ac} shows the data-collapse of $\frac{\mathrm{Var}(\ln(g))}{\expval{-\ln(g)}}$ for arm-chair ribbons of several widths, compared to the 1D chain described above.  
Numerically, the agreement is excellent, indicating that the scaling theory developed above applies equally to metallic armchair ribbons in this low-energy regime.

Finally, our scaling theory can be compared to the results of Ryu et. al. in \cite{RyuCrossover},  who derived FP-like equations   governing the evolution of the joint probability distribution of $x$ and the phase of the reflection coefficient $\phi$. 
These equations also predict a two-parameter scaling; however, the second parameter used in \cite{RyuCrossover} is proportional to:
\begin{equation}
	\tilde{r} = \frac{\epsilon}{V^2}.
\end{equation} 
For weak disorder $\tau \propto 1/ V^2$, and this differs from our scaling parameter $r = \epsilon\tau$ only by unimportant constant numerical factors.   However, for stronger disorder, when the Born approximation is no longer valid, the two scaling parameters are not equivalent.  Numerically, the data collapse with $\epsilon\tau$ is seen to work marginally better (see Appendix \ref{Sec:tilder}); however, the differences between the two parameters $r$ and $\tilde{r}$ are small for all values of $V$ considered here, such that the disagreement between the two is small.
A more stringent check of which scaling is most accurate would be  to consider systems with a non-linear dispersion, such as the zig-zag graphene ribbons; in such systems the energy dependence of the velocity leads to large differences between $r$ and $\tilde{r}$.  We defer such analysis to further work.

Ref. \cite{RyuCrossover}  also proposes an approximate solution describing the distribution in the cross-over regime.  In the limit $s \to \infty$, this becomes a Gaussian:
\begin{equation}
	P(x) = \frac{1}{\sqrt{\pi(1+\alpha)s}}\exp(-\frac{(x - (1-\alpha)s/2)^2}{(1+\alpha)s}),
	\label{eqn:ryu_asymptotic}
\end{equation}	
where $x\in \mathds{R}$ and $\alpha(\tilde{r})$ is a parameter that interpolates between the two limiting csaes. For $\alpha=1$ one recovers the distribution describing transport statistics in class BDI at zero energy (equation \ref{eqn:bdi_prob}), while for $\alpha = 0$ one obtains the distribution of class AI (equation \ref{eqn:ai_prob}) when $s\to\infty$.

This Gaussian form agrees qualitatively with equation \ref{eqn:gaussian_dsbn}. Away from zero energy, both distributions have negligible weight near $x=0$ when $s \to \infty$.  However, the two predictions differ quantitatively.  Comparing Eq. (\ref{eqn:ryu_asymptotic}) to Eq. (\ref{eqn:mu}), which is valid for $r \ll 1$, one finds that:
\begin{equation}
	\alpha = 1 - \frac{2}{\abs{\ln(r)}}.
\end{equation}
This is also what was found in \cite{RyuCrossover}. The variance of $x$ according to equation \ref{eqn:ryu_asymptotic} is then given by:
\begin{equation}
	\text{Var}(x) = \left(1 - \frac{1}{\abs{\ln(r)}} \right) s \approx s,
\end{equation}
when $r \ll 1$.  In contrast, numerically, $\text{Var}(x) \approx s/3$ (Eq. \ref{eqn:sigma}). This discrepancy cannot be fixed by changing some constants in equation \ref{eqn:ryu_asymptotic} since this would make the $\alpha = 0,1$ limits invalid.  Instead, the disagreement suggests that one of the assumptions used by Ref. \cite{RyuCrossover} to solve their differential equations is not justified in the case at hand.  This is not unexpected, since the authors of that work note that their approximate solution also disagrees with other approaches in certain limits.

\section{Conclusion} \label{Sec:Conc}

Transport statistics in the finite energy cross-over from BDI to AI cannot be described using the usual FP equation, which describes a 1-parameter scaling within each symmetry class. As such, alternative approaches are needed to study the cross-over in transport statistics from class BDI at zero energy to the large-energy limit where transport is expected to be well-described by class AI. The present work uses a numerical approach to study the phenomenology of this cross-over.

This approach yields three main insights.   First, the energy-dependence of the transport statistics away from zero energy is captured by introducing one additional parameter, $r = \epsilon\tau$, where $\tau$ is the scattering time.  This was demonstrated by means of the data collapse of several disorder averaged quantities over a wide range of parameters, obtained from the numerically computed conductance.  Such a 2-parameter scaling represents the minimal generalization of the FP equation needed to describe the cross-over regime. Notably, our scaling parameter differs slightly from that of previous works on the subject \cite{RyuCrossover}.

Second, transport can be phenomenologically divided into three main regimes.  For $r < 1$, when the wire is short compared to $ \xi_{\text{typ}}^2/l \sim \xi_{\text{av}}$, the system is not self-averaging, and the transport is well-approximated by that of the Chiral symmetry class.  As $s \rightarrow \infty$ for any $r >0$, the transport is governed by a Gaussian distribution with mean much larger than its standard deviation, and the resulting transport is in a localized regime. The cross-over regime interpolates between these two behaviors.

Third, after folding onto $x>0$, the transport statistics throughout the cross-over regime are well-approximated by a Nakagami distribution.  This distribution is described by 2 free parameters: a parameter $\gamma$ that sets the variance, and a parameter $\delta$ which controls the distribution's shape, which interpolates  continuously between the chiral ($\delta =1)$ and localized ($\delta \rightarrow \infty)$ regimes.    Physically, $\delta$ is proportional to the ratio $L/ \xi_{avg}$ of the length to the average localization length, and determines the degree to which the distribution is self-averaging. The distribution's  variance can be obtained from the numerical results, which yield Var$(x) \approx s/3$.  Interestingly, this disagrees with existing predictions of the cross-over.

This work raises several interesting questions about the nature of the cross-over from BDI to AI transport regimes in 1D.  First, if the cross-over regime merely reflects the nature of the disorder induced 1D critical point of class BDI, one would expect that  it exhibits universal features that are not specific to the single-channel case; as such it would be interesting to extend this study to the case of multi-channel wires.  A particularly interesting application of this would be to consider the scaling  of transport properties with width in metallic arm-chair graphene nanoribbons (for which the number of channels is always odd), which may shed insight on the behavior in the 2D limit. Second, it seems likely that the cross-over near the critical point in other 1D symmetry classes is described by the same scaling collapse with the $2$ parameters $s$ and $r$ used here.  It is even possible that the family of distributions identified here is super-universal, and captures transport in all of these cross-over regimes, along the lines of \cite{GruzbergSuperuniversality}.

\section{Acknowledgements}
The authors acknowledge useful conversations with Ilya Gruzberg and Xuzhe Ying. This work was supported primarily by the National Science Foundation through the University of Minnesota MRSEC under Award Number DMR-2011401. FJB acknowledges the financial support of NSF: DMR-1928166 and the Carnegie corporation of New York. AK was supported by the NSF grant DMR-2037654.

\appendix
\section{Statistical tests for class BDI} \label{Sec:StatTests}

A more rigorous test of the form of the distribution $P(x)$ can be performed using the methods of hypothesis testing. For instance, it was found that $P(x)$ is indistinguishable from the distribution of class BDI (equation \ref{eqn:bdi_prob}) when $L \ll \xi_{typ}^2/l$. To test this, one can take take the null hypothesis to be that $P(x)$ follows equation \ref{eqn:bdi_prob}, with $s$ obtained using a maximum likelihood estimator. The alternative hypothesis is that $P(x)$ does not follow this distribution. One can then perform a statistical test to check if the null hypotheis can be rejected. In this case, the Kolmogorov-Smirnov test is used. This test uses the maximum distance between the empirical cumulative distribution function (CDF) and the expected CDF using \ref{eqn:bdi_prob} as the test statistic. One then obtains a p-value, which is the probability of obtaining a test statistic at least as extreme as the one observed, assuming the null hypothesis is true. If the p-value is less than a pre-determined value $\alpha$, one can reject the null hypothesis. An observed p-value less than $\alpha$ means that the probability of rejecting the null hypothesis if it is true, known as the type-I error rate, is at most $\alpha$. Here $\alpha$ is taken to be $0.05$, a value used commonly. Note that if a p-value greater than $\alpha$ is obtained, one cannot confirm the null hypothesis.

\begin{figure}
	\centering
	\includegraphics[width=\linewidth]{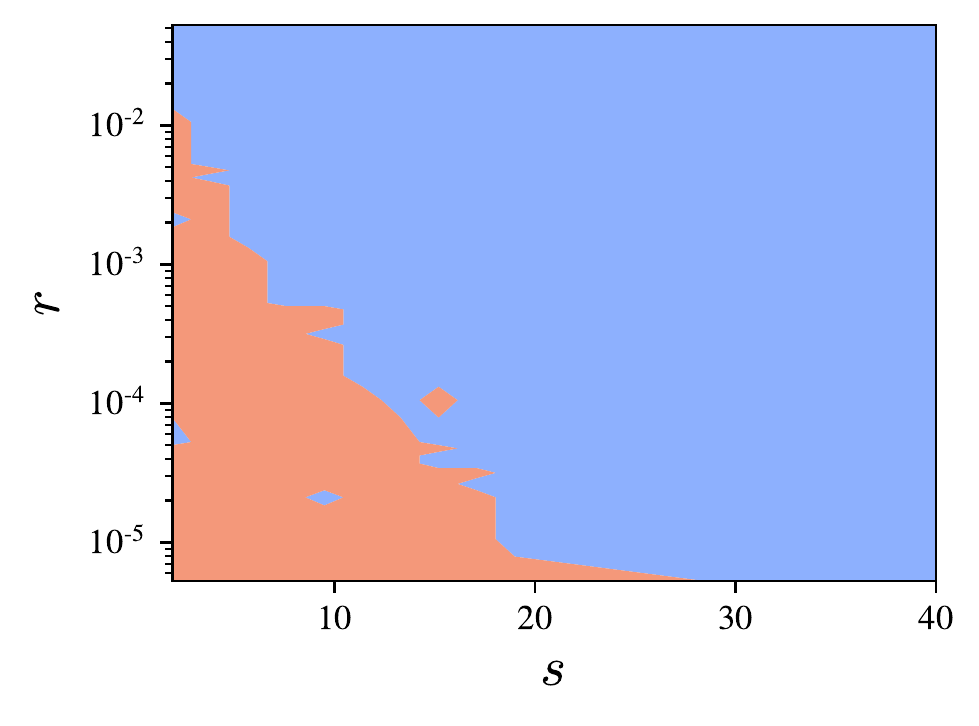}
	\caption{The results of the Kolmogorov-Smirnov test of $P(x)$ with the null hypothesis that $P(x)$ follows equation \ref{eqn:bdi_prob}, shown as a function of the scaling parameters $r$ and $s$. $V=0.5$ is used here. In the blue region, the p-value is less than $\alpha=0.05$, therefore the distribution is not identical to that of class BDI here. In the red region, the p-value is greater than $\alpha$, so $P(x)$ is not inconsistent with that of class BDI.}
	\label{fig:bdi_p_value}
\end{figure}

All of this is performed using methods available in the SciPy package in python \cite{SciPy}. The results of the test are shown in figure \ref{fig:bdi_p_value}. One finds that $P(x)$ is indistinguishable from that of class BDI up to a value of $s$ that increases when $r$ decreases.

\section{Gaussian limit of equation \ref{eqn:gamma_dsbn}} \label{GaussApp}

The distribution of equation \ref{eqn:gamma_dsbn} becomes a Gaussian distribution in the limit $\delta \to \infty$. 
The moments of the distribution are then given by:
\begin{equation}
	\mu_n = \expval{x^n} =\gamma^n \frac{\Gamma\left(\frac{\delta+n}{2}\right)}{\Gamma\left(\frac{\delta}{2}\right)}.
\end{equation}
The mean (first moment) in the limit $\delta \to \infty$ can be found using:
\begin{equation}
	\lim_{\delta\to \infty} \frac{\Gamma\left(\frac{\delta+1}{2}\right)}{\Gamma\left(\frac{\delta}{2}\right)}  = \sqrt{\frac{\delta}{2}}.
\end{equation}
The mean of the distribution is therefore $\mu =\gamma \sqrt{\delta/2}$. Next, one can compute the limits of the higher cumulants. For the second cumulant (variance) one finds:
\begin{equation}
	\sigma^2 = \gamma^2 \lim_{\delta\to \infty} \left[ \frac{\Gamma\left(\frac{\delta+2}{2}\right)}{\Gamma\left(\frac{\delta}{2}\right)} - \left( \frac{\Gamma\left(\frac{\delta+1}{2}\right)}{\Gamma\left(\frac{\delta}{2}\right)} \right)^2 \right] = \frac{\gamma^2}{4}.
\end{equation}
This establishes the results quoted in equation \ref{eqn:gamma_dsbn_limit}. 
The higher cumulants vanish at least as fast as $1/\sqrt{\delta}$, indicating that  in the limit $\delta \to \infty$ this distribution converges to a Gaussian. 
To see this, note that as $\delta \to \infty$ the distribution is well-approximated by
\begin{align}
	\begin{split}
		P(x) & = \frac{2}{\gamma \Gamma(\delta/2 ) } \text{exp} \Bigg[ - \left( \frac{x}{\gamma}\right)^2 \\
		& - ( \delta - 1)  \log \frac{ \bar{x} }{ \gamma}
		+  2\frac{\bar{x} \tilde{x}}{\gamma^2}  - \left( \frac{ \tilde{x}}{\gamma}\right)^2  \Bigg]
	\end{split}
\end{align}
where $\bar{x} = \sqrt{\frac{\delta - 1}{2}}$ is the value where the distribution is maximal, $\tilde{ x} = (x - \bar{x} )a$, and the terms that have been dropped in the exponential are of order $\frac{\tilde{x}^3}{\sqrt{\delta - 1}}$.  It follows that up to corrections that vanish like $\frac{1}{\sqrt{\delta}}$, the distribution is Gaussian.  

\section{Difference between the scaling parameters $r$ and $\tilde{r}$} \label{Sec:tilder}

\begin{figure}
	\centering
	\includegraphics[width=\linewidth]{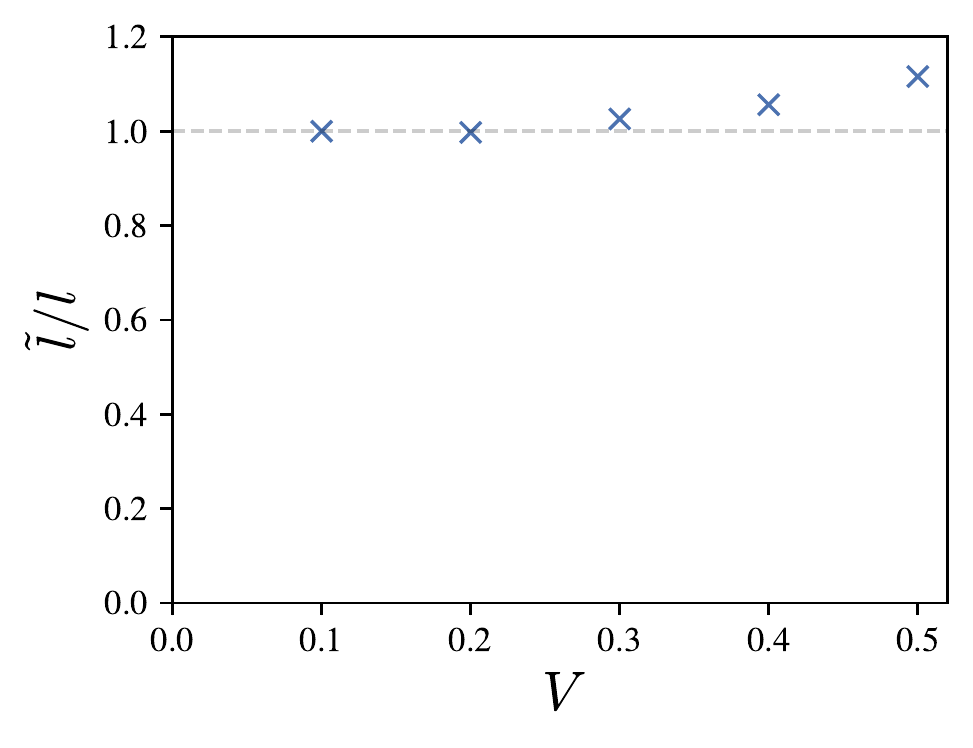}
	\caption{The ratio of the scaled mean free path and the actual mean free path as a function of the disorder strength $V$, showing that $1/V^2$ scaling is no longer accurate when the disorder is strong. $\tilde{l}$ is given by $l_0 \left(\frac{0.1}{V}\right)^2$, where $l_0$ is the mean-free path for $V=0.1$.}
	\label{fig:l_comparison}
\end{figure}

Two-parameter scaling was also predicted in \cite{RyuCrossover}.  However, the second scaling parameter predicted by their treatment is  $\tilde{r} = \epsilon/V^2$, where $V$ is the disorder strength.   In the weak disorder limit where the Born approximation is applicable this is equivalent to the scaling parameter $r = \epsilon\tau$ used here; however, the two differ for larger values of $V$.   
This is apparent in Figure \ref{fig:l_comparison}, which  compares the exact numerical value of the mean free path to the Born approximation result $1/V^2$; the difference between the two increases as the disorder becomes stronger.   However, the deviations are small, even for the largest  disorder strengths considered here.   This discrepancy translates to a similar difference in the lifetime $\tau$. 

\begin{figure}
	\centering
	\includegraphics[width=\linewidth]{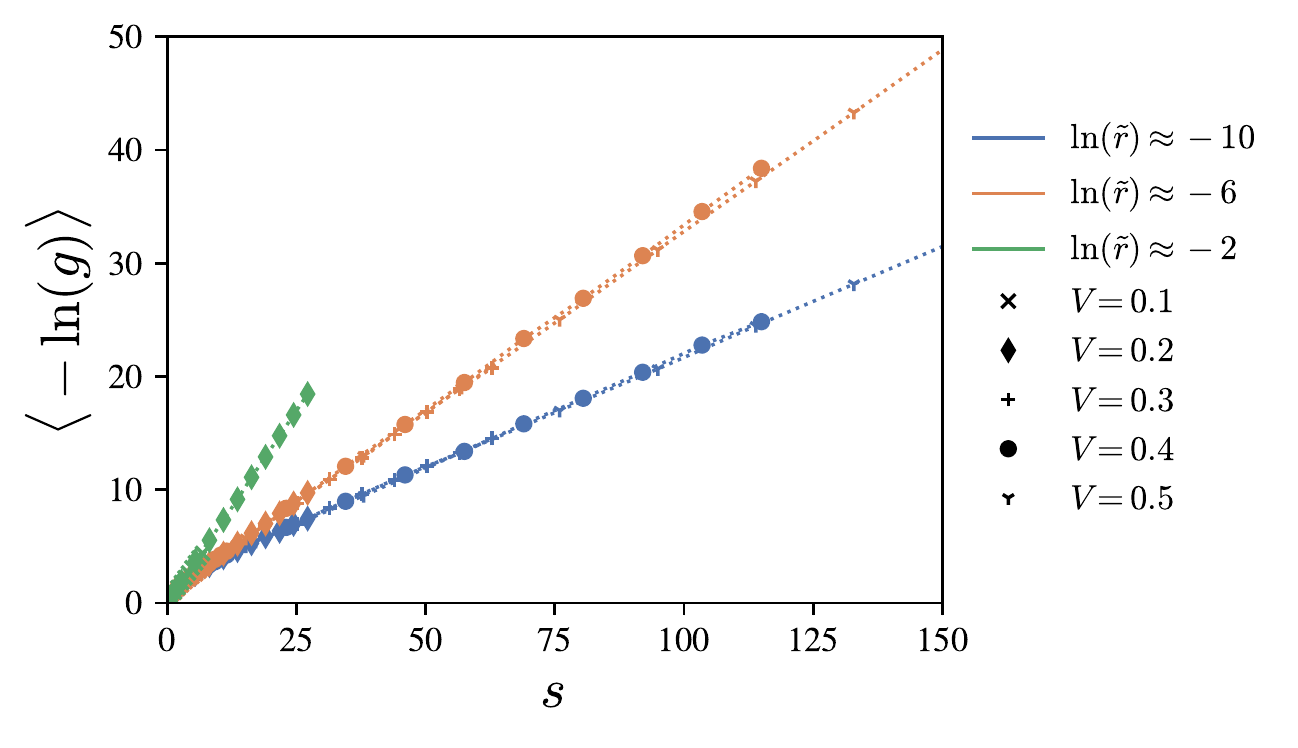}
	\caption{Same as figure \ref{fig:lng_collapse}, but using $\tilde{r}$ as the second parameter. }
	\label{fig:lng_collapse_2}
\end{figure}

Figure \ref{fig:lng_collapse_2} shows the data collapse using $\tilde{r}$ rather than $r$ as the second scaling variable.  Since the two parameters differ by at most $10 \%$, the collapse still works fairly well; however the collapse is noticeably worse than in Fig. \ref{fig:lng_collapse}.  This suggests that  the FP equations of \cite{RyuCrossover}, which predict scaling collapse with $\tilde{r}$, rely on approximations that break down for large $V$.  Notably, they are obtained when $l$ is much larger than the lattice spacing. Since $l \approx 5$ when $V=0.5$, this assumption is not well justified at the strongest disorder strengths studied here, where the two scalings differ.  

\bibliographystyle{apsrev4-2}
\bibliography{paper_references}

\end{document}